\documentclass[notitlepage,aps,prd,preprintnumbers,nofootinbib,longbibliography,showpacs]{revtex4-1}
\usepackage[%
  colorlinks=true,
  urlcolor=blue,
  linkcolor=blue,
  citecolor=blue
]{hyperref}
\usepackage{slashed,color,amsmath,amssymb,mathrsfs}
\usepackage{graphicx}
\usepackage{hyperref}
\usepackage{longtable}
\usepackage{array}
\usepackage{accents}

\allowdisplaybreaks

\newcommand{\la}{\langle}
\newcommand{\ra}{\rangle}
\newcommand{\chip}{\chi_+}
\newcommand{\chim}{\chi_-}
\newcommand{\fp}{f_+}

\newcommand{\ud}{u^\dag}

\newcommand{\Gam}{\Gamma}
\newcommand{\gamf}{\gamma_5}
\newcommand{\fgamma}{\gamf\gamma}
\newcommand{\psib}{\bar\psi}

\begin{document}
\title{Chiral Lagrangians with $\Delta(1232)$ to one loop}
\author{Shao-Zhou Jiang$^{1,2}$}\email{jsz@gxu.edu.cn}
\author{Yan-Rui Liu$^3$}\email{yrliu@sdu.edu.cn}
\author{Hong-Qian Wang$^1$}
\affiliation{$^1$Department of Physics and GXU-NAOC Center for Astrophysics and Space Sciences, Guangxi University,
Nanning, Guangxi 530004, People's Republic of China\\
$^2$Guangxi Key Laboratory for the Relativistic Astrophysics, Nanning, Guangxi 530004, People's Republic of China\\
$^3$School of Physics and Key Laboratory of Particle Physics and Particle Irradiation (MOE),
Shandong University, Jinan 250100, People's Republic of China}

\begin{abstract}
We construct the Lorentz-invariant chiral Lagrangians up to the order $\mathcal{O}(p^4)$ by including $\Delta(1232)$ as an explicit degree of freedom. A full one-loop investigation on processes involving $\Delta(1232)$ can be performed with them. For the $\pi\Delta\Delta$ Lagrangian, one obtains 38 independent terms at the order $\mathcal{O}(p^3)$ and 318 independent terms at the order $\mathcal{O}(p^4)$. For the $\pi N\Delta$ Lagrangian, we get 33 independent terms at the order $\mathcal{O}(p^3)$ and 218 independent terms at the order $\mathcal{O}(p^4)$. The heavy baryon projection is also briefly discussed.
\end{abstract}
\maketitle

\section{Introduction}\label{intr}

The lowest excited states of the nucleon are the four $\Delta(1232)$ baryons, which play an important role in the low-energy processes, such as the $\pi N$ scattering, the magnetism of the nucleon, the electromagnetic interactions of nucleons and so on, because of the strong coupling between the $\Delta$ and the nucleon. Their mass gap is around 300 MeV which is not a big number and they may be treated as degenerate states in the large $N_c$ limit \cite{tHooft:1973alw,Witten:1979kh}. At present, because of the difficulties in solving the nonperturbative QCD problem, chiral perturbation theory (ChPT) \cite{weinberg,GS1,GS2,Gasser:1987rb} is still a feasible and efficient method to describe the low-energy processes involving pions and nucleons. Due to the above reasons, the effects of the $\Delta$ baryons in ChPT are worth separated from low energy constants (LECs) and one can included $\Delta$ baryons in this framework as explicit degrees of freedom \cite{Jenkins:1991es,Hemmert:1997ye}.

To include $\Delta$ in ChPT, one usually needs to set up an expansion method, i.e. a power counting scheme, because a new scale larger than the nucleon mass appears. In the pion meson sector, the Lagrangian and the S-matrix are expanded with the power of meson mass (or its energy-momentum) over the scale of chiral symmetry breaking, $p/\Lambda_\chi\sim m_\pi/\Lambda_\chi$ with $\Lambda_\chi\sim 1$ GeV. When the nucleon is included in ChPT, because its mass $m_N$ is comparable to $\Lambda_\chi$, this simple power counting becomes problematic. In the literature, there are efforts to solve this problem with heavy baryon formalism \cite{Jenkins:1990jv}, infrared regularization \cite{Becher:1999he}, or EOMS regularization scheme \cite{Fuchs:2003qc}. After the introduction of the scale $m_\Delta$, a new expansion parameter $\delta/\Lambda_\chi=(m_\Delta-m_N)/\Lambda_\chi$ is involved. One may choose a power counting scheme, $m_\pi/\Lambda_\chi\sim\delta/\Lambda_\chi$ \cite{Hemmert:1997ye} or $m_\pi/\Lambda_\chi\sim(\delta/\Lambda_\chi)^2$ \cite{Pascalutsa:2002pi}, to calculate the S-matrix. Much deeper understanding about the power counting problems in ChPT with explicit $\Delta$ could be obtained in the future once high order Lagrangians are given. At present, the full chiral Lagrangian with $\Delta$ is still at low orders and we would like to construct high order Lagrangians in this paper. Here we will adopt the small scale expansion scheme \cite{Hemmert:1997ye} and simply use $p$ to denote $m_\pi$, $\delta$, or the three momentum of the nucleon or $\Delta$.

Up to now, the chiral Lagrangians for mesons have been obtained up to the ${\mathcal O}(p^6)$ order (two-loop level) \cite{GS1,GS2,p61,p62,p6p,p6a1,p6a2,tensor1,U3,ourf}, including the whole 16 bilinear light-quark currents (scalar, pseudoscalar, vector, axial-vector, and tensor) of the special unitary group and the unitary group. For baryon ChPT, the full Lagrangians up to the order ${\mathcal O}(p^4)$ (one-loop level) were completed recently \cite{Gasser:1987rb,Krause:1990xc,Ecker:1994pi,1998NuPhA.640..199F,Meissner:1998rw,pin4,pib31,pib32,Jiang:2016vax}. For ChPT with $\Delta$, the Lagrangian is only at the order ${\mathcal O}(p^2)$ (tree level) \cite{Hemmert:1997ye}. In principle, high order calculations of the S-matrix in ChPT will improve the precision of the theory. Specific processes using part of high order terms have been studied, such as the nucleon-$\Delta$ transition \cite{Pascalutsa:1999zz,Pascalutsa:2005ts,Procura:2008ze,Li:2017vmq}, the properties of $\Delta$ \cite{Gellas:1998wx,Pascalutsa:2004je,Pascalutsa:2006up,Pascalutsa:2007yg,Pascalutsa:2007wb,Bernard:2007cm,Bernard:2009mw,Ren:2013oaa},  scattering processes with nucleon or $\Delta$ \cite{Fettes:2000bb,Hildebrandt:2005ix,Bernard:2005fy,Pascalutsa:2005vq,Liu:2007ct,Liu:2010bw,Alarcon:2012kn,McGovern:2012ew,Yao:2016vbz,HillerBlin:2016jpb}, and so on. Completing high order Lagrangians with $\Delta$ is the first task in improving the precision of ChPT and will be helpful to deeper understanding of the chiral expansion in various processes. In this paper, we would like to construct the chiral Lagrangians with $\Delta$ up to the $\mathcal{O}(p^4)$ order (one-loop level).

This work is organized as follows. In Sec. \ref{freef}, we provide the basic properties of spin-$\frac{3}{2}$ fields in the Rarita-Schwinger formalism. It is the base for the discussion about $\Delta$. In Sec. \ref{rev}, we review the building blocks for the construction of the chiral Lagrangians without $\Delta$. Some properties also work in the $\Delta$ case. In Sec. \ref{rev1}, we present the building blocks used in constructing chiral Lagrangians with $\Delta$. In Sec. \ref{const}, the properties of the building blocks are given. With these properties, a systematic method for the construction of Lagrangians is introduced. In Sec. \ref{results}, we list our results and present some discussions. Section \ref{summ} is a short summary.

\section{Spin-$\frac{3}{2}$ fields and their basic properties}\label{freef}

$\Delta(1232)$ is a spin-$\frac{3}{2}$ field. There are many ways to describe such a field \cite{Dirac:1936tg,Fierz:1939ix,Rarita:1941mf,1954PhRv...95.1334G}. In this paper, we adopt the widely used vector-spinor representation $\Psi^\mu$ ($\mu=0,1,2,3$) \cite{Rarita:1941mf} to give the Lagrangian. However, the vector-spinor representation (Rarita-Schwinger or RS field) contains two unphysical spin-$\frac{1}{2}$ degrees of freedom and one needs extra conditions to restrict the representation. Because of the unphysical spin-$\frac{1}{2}$ components, an arbitrary unphysical parameter $A$ exists in the Lagrangian. Before the discussion of $\Delta$ in ChPT, we first review some basic properties for the free RS field. More details can be found in Refs. \cite{Dirac:1936tg,Fierz:1939ix,Rarita:1941mf,1954PhRv...95.1334G,Moldauer:1956zz,1958NCim....9S.416F,Aurilia:1969bg,VanNieuwenhuizen:1981ae,Williams:1985zz,Benmerrouche:1989uc,Pascalutsa:1994tp,Haberzettl:1998rw,Pilling:2004cu,2006PAN....69..541K}.

The general Lagrangian for the free RS field with mass $m_\Delta$ is \cite{Moldauer:1956zz}
\begin{eqnarray}\label{rsl}
\mathscr{L}_{\mathrm{f}}&=&\bar{\Psi}_{\mu}\Lambda_A^{\mu\nu}\Psi_\nu,\\
\Lambda_A^{\mu\nu}&=&-\big[(i\slashed{\partial}-m_\Delta)g^{\mu\nu}+iA(\gamma^\mu\partial^\nu+\gamma^\nu\partial^\mu)\nonumber\\
&&+\frac{i}{2}(3A^2+2A+1)\gamma^\mu\slashed{\partial}\gamma^\nu\nonumber\\
&&+m_\Delta(3A^2+3A+1)\gamma^\mu\gamma^\nu\big],\nonumber
\end{eqnarray}
where $A\neq-1/2$ is an arbitrary real number. Generally speaking, $A$ can also be a complex number. In this case, the Lagrangian needs some modifications \cite{1958NCim....9S.416F,Pilling:2004cu},
\begin{eqnarray}
\Lambda_A^{\mu\nu}&=&-\big[(i\slashed{\partial}-m_\Delta)g^{\mu\nu}+i(A\gamma^\mu\partial^\nu+A^*\gamma^\nu\partial^\mu)\notag\\
&&+\frac{i}{2}(3AA^*+A+A^*+1)\gamma^\mu\slashed{\partial}\gamma^\nu\nonumber\\
&&+m_\Delta(3AA^*+\frac{3}{2}A+\frac{3}{2}A^*+1)\gamma^\mu\gamma^\nu\big].
\end{eqnarray}
Here, an overall minus sign has been chosen \cite{Tang:1996sq,Hemmert:1997ye,Hacker:2005fh,Krebs:2009bf,Scherer:2012xha}. It ensures that the spatial components of $\Psi^\mu$ behave like a Dirac field and the Hamiltonian is positive definite \cite{Tang:1996sq}. In the following, $A$ is considered to be real. A complex $A$ would only result in complicated expressions.

The above Lagrangian is invariant under the so called ``point" or ``contact" transformation,
\begin{align}
\Psi_\mu&\to\Psi'_\mu=\Psi_\mu+\frac{1}{2}a\gamma_\mu\gamma_\nu\Psi^\nu,\label{pt1}\\
A&\to A'=\frac{A-a}{1+2a},\quad a\neq-\frac{1}{2},\label{pt2}
\end{align}
which is not a symmetry of the Lagrangian because the parameter $A$ is changed. The choice for the value of $A$ does not affect physical quantities \cite{Kamefuchi:1961sb,Pilling:2004cu,Krebs:2009bf}. In studying various physical processes involving the RS field, one can choose any suitable $A$ for the purpose of convenient use. If $A=-\frac{1}{3}$, the original Rarita-Schwinger Lagrangian is recovered \cite{Rarita:1941mf}. If $A=-1$, the propagator has a very simple and widely used form \cite{Benmerrouche:1989uc}.

To restrict arbitrariness and to simplify the Lagrangian, one may adopt the method proposed by Pascalutsa in Ref. \cite{Pascalutsa:1994tp} where a point invariant RS field $\psi_A^\mu\equiv O_A^{\mu\nu}\Psi_{\nu}=(g^{\mu\nu}+\frac{1}{2}A\gamma^\mu\gamma^\nu)\Psi_{\nu}$ is defined. Then the new form of the Lagrangian reads
\begin{eqnarray}
\mathscr{L}_{\mathrm{f}}&=&\bar{\psi}_{A\mu}\Lambda^{\mu\nu}\psi_{A\nu},\\
\Lambda^{\mu\nu}&=&-(i\slashed{\partial}-m_\Delta)g^{\mu\nu}+\frac{1}{4}\gamma^\mu\gamma^\lambda(i\slashed{\partial}-m_\Delta)\gamma_\lambda\gamma^\nu.\nonumber
\end{eqnarray}
Now, all the variations to $A$ are implicitly contained in $\psi_{A}^\mu$ and $\Lambda^{\mu\nu}$ is independent of $A$.

With the Euler-Lagrange equation and some techniques such as that given in the appendix of Ref. \cite{Nath:1971wp}, one obtains
\begin{align}
&(i\slashed{\partial}-m_\Delta)\Psi_\mu=0,\label{eompsif}\\
&\gamma^\mu\Psi_\mu=0,\label{eoms1}\\
&\partial^\mu\Psi_\mu=0.\label{eoms2}
\end{align}
Eq. \eqref{eompsif} is the equation of motion for the RS field and Eqs. \eqref{eoms1} and \eqref{eoms2} are two subsidiary conditions to eliminate the redundant components of the RS field. The redefined field $\psi_A^\mu$ also satisfies these equations.

When one considers interactions of $\Delta$, e.g. $\Delta\Delta\gamma$ or $\Delta N\pi$ interaction, a covariant derivative is needed in the Lagrangian and one more free parameter ($z$ parameter) also appears. To be consistent with the point transformation, this $z$ parameter is necessary and its value can be obtained from experiments \cite{Nath:1971wp}.

The framework to include $\Delta$ in ChPT can be found in Ref. \cite{Tang:1996sq}. It gives the method to separate out the redundant degrees of freedom and reveal the physical degrees of freedom explicitly. Later, a series of systematic works analyze the structures of chiral Lagrangian with $\pi$ and $\Delta$ fields and give the leading order Lagrangian \cite{Hemmert:1996xg,Hemmert:1997wz,Hemmert:1997ye,Hacker:2005fh}. In Ref. \cite{Krebs:2009bf}, the off-shell parameters in ChPT with $\Delta$ are proved unphysical and can be removed. In the following parts, we consider systematically the structures of chiral Lagrangian with $\Delta$ and construct the Lagrangian to the one-loop order by eliminating the redundant degrees of freedom.

\section{Building blocks in constructing chiral Lagrangians}

In ChPT, the Lagrangian is invariant with respect to various QCD symmetries. We need to know the transformations of various building blocks (baryon or meson fields, external sources, or combined structures of them). In this section, we review briefly the building blocks of chiral Lagrangian without $\Delta$ and present new building blocks involving $\Delta$. The details for the former case can be found in Refs. \cite{GS1,GS2,p61,p62,p6a2,ourf,pin4,pib31,Jiang:2016vax}.

\subsection{Building blocks without $\Delta$}\label{rev}

The two-flavor QCD Lagrangian $\mathscr{L}$ can be written as
\begin{eqnarray}
\mathscr{L}=\mathscr{L}^0_{\mathrm{QCD}}+\bar{q}(\slashed{v}+\slashed{a}\gamma_5-s+ip\gamma_5)q~,
\end{eqnarray}
where $\mathscr{L}^0_{\mathrm{QCD}}$ is the original QCD Lagrangian and $q$ is the quark field $u$ or $d$. We use $s$, $p$, $v^\mu$, and $a^\mu$ to denote scalar, pseudoscalar, vector, and axial-vector external sources, respectively. Conventionally, the tensor source and the $\theta$ term are ignored. In ChPT, $a^\mu$ is usually traceless. If we use $S$ to denote the external source $s$, $p$, $v^\mu$, or $a^\mu$, one can divide it into a traceless part $S^i$ ($i=1,2,3$) and a trace part $S_s$
\begin{eqnarray}\label{source}
&S=S^i\tau_i+S_sI_2,\\
&S^i=\frac{1}{2}\la S\tau^i\ra,~ S_s=\frac{1}{2}\la S\ra,\nonumber
\end{eqnarray}
where $\tau^i$ are Pauli matrices, $I_2$ is the $2\times2$ identity matrix, and $\la\cdots\ra$ denotes the trace of ``$\cdots$'' in isospin space. For convenience, we here use the convention $S^i\tau_i=S^i\tau^i=S_i\tau_i$. In the following, we will perform similar decompositions for the $2\times 2$ matrices in the isospin space.

The QCD Lagrangian $\mathscr{L}^0_{\mathrm{QCD}}$ exhibits a global $SU(2)_L\times SU(2)_R$ chiral symmetry when the light quarks are massless. This symmetry is spontaneously broken into $SU(2)_V$ and three Goldstone bosons (pseudoscalar mesons) appear. These pseudoscalar mesons get their masses once the light quark masses are considered. In ChPT, these mesons are collected into an $SU(2)$ matrix $U$ whose transformation is $U\to g_L U g_R^\dag$. Here $g_L$ and $g_R$ represent $SU(2)_L$ and $SU(2)_R$ chiral rotations, respectively. Usually, another field $u$ is defined through $u^2=U$. It transforms as $u\to g_L uh^\dag=h ug_R^\dag$ under the chiral rotation, where $h$ is a compensator field and a function of the pion fields.

To construct the chirally invariant Lagrangian, one introduces several combinations of the external sources and meson fields. Such structures, called building blocks, are
\begin{align}\label{df}
\begin{split}
u^\mu&=i\{u^\dag(\partial^\mu-ir^\mu)u-u(\partial^\mu-il^\mu)u^\dag\},\\
\chi_\pm&=u^\dag\chi u^\dag\pm u\chi^\dag u,\\
h^{\mu\nu}&=\nabla^\mu u^\nu+\nabla^\nu u^\mu,\\
f_+^{\mu\nu}&=u F_L^{\mu\nu} u^\dag+ u^\dag F_R^{\mu\nu} u,\\
f_-^{\mu\nu}&=u F_L^{\mu\nu} u^\dag- u^\dag F_R^{\mu\nu} u=-\nabla^\mu u^\nu+\nabla^\nu u^\mu,
\end{split}
\end{align}
where $r^\mu=v^\mu+a^\mu$, $l^\mu=v^\mu-a^\mu$, $\chi=2B_0(s+ip)$, $F_R^{\mu\nu}=\partial^{\mu}r^{\nu}-\partial^{\nu}r^{\mu}-i[r^{\mu},r^{\nu}]$, $F_L^{\mu\nu}=\partial^{\mu}l^{\nu}-\partial^{\nu}l^{\mu}-i[l^{\mu},l^{\nu}]$, and $B_0$ is a constant related to the quark condensate. The definition of the covariant derivative $\nabla^{\mu}$ acting on any building block $X$ in Eq. (\ref{df}) is
\begin{eqnarray}\label{cd}
&\nabla^{\mu}X\equiv\partial^\mu X+[\Gamma^\mu,X],&\\
&\Gamma^{\mu}=\frac{1}{2}\{\ud(\partial^\mu-ir^\mu)u+u(\partial^\mu-il^{\mu})\ud\}.&\nonumber
\end{eqnarray}
In constructing the Lagrangian, the following two relations will be useful
\begin{eqnarray}
&[\nabla^\mu,\nabla^\nu]X=[\Gamma^{\mu\nu},X],\label{Gam0}&\\
&\Gamma^{\mu\nu}=\nabla^{\mu}\Gam^{\nu}-\nabla^{\nu}\Gam^{\mu}-[\Gam^{\mu},\Gam^{\nu}]
=\frac{1}{4}[u^\mu,u^\nu]-\frac{i}{2}f_{+}^{\mu\nu}.\label{Gam1}&
\end{eqnarray}

Besides the above meson and external fields, we also need baryons. In the following parts, the nucleon doublet is denoted as
\begin{eqnarray}
\psi=\begin{pmatrix}
p\\n
\end{pmatrix}
\end{eqnarray}
and its covariant derivative is defined as $D^\mu\psi\equiv(\partial^\mu+\Gamma^{\mu})\psi$. Note that we use a different symbol for the covariant derivative acting on baryons in this paper.

\subsection{New building blocks involving $\Delta$}\label{rev1}
Unlike the nucleon, it is a bit complex to describe $\Delta$ fields in the isospin space. In the literature, they are usually denoted by an isovector-isospinor $\psi^\mu_i$ \cite{Tang:1996sq,Hemmert:1996xg,Hemmert:1997wz,Hemmert:1997ye,Hacker:2005fh,Krebs:2009bf,Scherer:2012xha}. To eliminate the two redundant isospin-$1/2$ degrees of freedom, a subsidiary condition is imposed
\begin{eqnarray}
\tau^i\psi^\mu_i=0\qquad (i=1,2,3).\label{sctau}
\end{eqnarray}
The representation for the $I=\frac32$ components is
\begin{align}
\psi^\mu_1&
=\frac{1}{\sqrt{2}}\begin{pmatrix}
\frac{1}{\sqrt{3}}\Delta^{0}-\Delta^{++}\\
\Delta^--\frac{1}{\sqrt{3}}\Delta^+
\end{pmatrix}^\mu,\nonumber\\
\psi^\mu_2&
=-\frac{i}{\sqrt{2}}\begin{pmatrix}
\frac{1}{\sqrt{3}}\Delta^{0}+\Delta^{++}\\
\Delta^-+\frac{1}{\sqrt{3}}\Delta^+
\end{pmatrix}^\mu,\nonumber\\
\psi^\mu_3&
=\sqrt{\frac{2}{3}}\begin{pmatrix}
\Delta^{+}\\
\Delta^0
\end{pmatrix}^\mu,
\end{align}
where the signs are the same as those in Refs. \cite{Tang:1996sq,Scherer:2012xha}. A different overall sign was adopted in Ref. \cite{Hemmert:1997ye}.

Note that each $\psi_i^\mu$ is an isospin doublet and the index $i$ needs to be contracted with the isovector index of another field. Thus we need to reveal the implicit isospin indices of the building blocks in Eq. \eqref{df}. Similar to Eq. (\ref{source}), we decompose each building block $X$ ($u^\mu$, $h^{\mu\nu}$, $f_{\pm}^{\mu\nu}$, or $\chi_{\pm}$) with the following formulas
\begin{eqnarray}\label{Xi}
&X=X_i\tau_i+X_sI_2,&\\
&X_i=\frac{1}{2}\la X\tau_i\ra,\quad X_s=\frac{1}{2}\la X\ra.&\nonumber
\end{eqnarray}
Specifically, one has
\begin{eqnarray}
f_{+,s}^{\mu\nu}&=&2\partial^\mu v_s^\nu-2\partial^\nu v_s^\mu,
\end{eqnarray}
which is related to the external source $v^\mu$ only.

When we consider only the $\pi\Delta\Delta$ interactions, the chiral Lagrangian is the linear combination of $\psib_{i}^\mu {\cal O}^{ij}_{A\mu\nu}\psi_{j}^{\nu}$ where ${\cal O}^{ij}_{A\mu\nu}$ containing the pion fields and external sources with various Lorentz structures depends on $A$. The LECs in the Lagrangian are also $A$-dependent. For the $\pi N\Delta$ interactions, some extra $z_n$ parameters are needed in the Lagrangian because of the point transformation \cite{Nath:1971wp}. The interaction terms have the form $\psib {\cal O}^{i\mu}\Theta_{A,n,\mu\nu}(z_n)\psi_{i}^{\nu}+\mathrm{h.c.}$, where ${\cal O}^{i\mu}$ containing the pion fields and external sources with various Lorentz structures is independent of $A$ and
\begin{eqnarray}\label{z-para}
\Theta_{A,n,\mu\nu}(z_n)&=&g_{\mu\nu}+[z_n+\frac{1}{2}(1+4z_n)A]\gamma_\mu\gamma_\nu\nonumber\\
&=&(g_{\mu\alpha}+z_n\gamma_\mu\gamma_\alpha)({g^\alpha}_\nu+\frac12A\gamma^\alpha\gamma_\nu)\nonumber\\
&=&(g^\alpha_\mu+\frac12A\gamma_\mu\gamma^\alpha)(g_{\alpha\nu}+z_n\gamma_\alpha\gamma_\nu)\nonumber\\
&\equiv&\Theta_{n,\mu\alpha}(z_n){O_{A}^\alpha}_\nu={O_{A\mu}}^\alpha\Theta_{n,\alpha\nu}(z_n).
\end{eqnarray}
Here, $n$ denotes the order of a term and $z_n$ is a free parameter which can be obtained from experiments. We do not discuss how to determine $z_n$ in this work. The LECs in the $\pi N\Delta$ Lagrangian are independent of $A$.

One can absorb the arbitrary parameter $A$ into the redefinition of the RS field (as in the free RS case) and use $\psi_{Ai}^{\mu}\equiv(g^{\mu\nu}+\frac{1}{2}A\gamma^\mu\gamma^\nu)\psi_{i\nu}=O_{A}^{\mu\nu}\psi_{i\nu}$ and $\psi_{A,n,i}^{\mu}\equiv\Theta_{n}^{\mu\nu}(z_n)\psi_{Ai\nu}\equiv O_A^{\mu\nu}\psi_{n,i,\nu}$ to construct chiral Lagrangian. In this case, the $\pi\Delta\Delta$ interaction terms have the form $\psib_{Ai}^\mu {\cal O}^{ij}_{\mu\nu}\psi_{Aj}^{\nu}$ where ${\cal O}^{ij}_{\mu\nu}$ and the LECs are independent of $A$, and the $\pi N\Delta$ interaction terms have the form $\psib {\cal O}_{i\mu}\psi_{A,n,i}^\mu+\mathrm{h.c.}$.\\

To summarize, the building blocks in constructing chiral Lagrangians with $\Delta$ are all $X_i$ and $X_s$ in Eq. \eqref{Xi}, the Dirac fields $\psi$ and $\psib$, the RS fields $\psi_{i}^{\mu}$, $\psib_{i}^{\mu}$, $\psi_{A,n,i}^{\mu}$, and $\psib_{A,n,i}^{\mu}$, and their covariant derivatives. The covariant derivatives of these building blocks are
\begin{eqnarray}
\nabla^\mu X^i&=&\partial^\mu X^i-2i\epsilon^{ijk}X_j\Gamma_k^{\mu},\nonumber\\
\nabla^\mu X_s&=&\partial^\mu X_s,\nonumber\\
D^\mu\psi&=&\partial^\mu\psi+(\Gamma_i^{\mu}\tau^i+\Gamma_s^{\mu}I_2)\psi,\nonumber\\
D^\mu\psi^{i\nu}&=&\partial^\mu\psi^{i\nu}-2i\epsilon^{ijk}\Gamma_k^{\mu}\psi_{j}^{\nu}+\Gamma_j^{\mu}\tau^j\psi^{i\nu}+\Gamma_s^{\mu}\psi^{i\nu},
\end{eqnarray}
where $\Gamma^\mu=\Gamma^\mu_i\tau^i+\Gamma_s^\mu I_2$. As an alternative choice, in the $\pi\Delta\Delta$ case, one may adopt the redefined RS fields $\psi_{Ai}^\mu$ and $\psib_{Ai}^{\mu}$ (instead of $\psi_{i}^{\mu}$ and $\psib_{i}^{\mu}$) and their covariant derivatives to construct Lagrangians. We will discuss both cases later.

\section{Construction of chiral Lagrangians with $\Delta$}\label{const}

In this section, we introduce the method to construct chiral Lagrangians with $\Delta$ step by step. The construction procedure in this method is very similar to that used to construct the meson and meson-baryon chiral Lagrangians in Refs. \cite{ourf,Jiang:2016vax}.

\subsection{Power counting and transformation properties}\label{bbp}

We adopt the chiral dimensions for the building blocks assigned in Refs. \cite{GS1,GS2,Gasser:1987rb,p62,pin4,pib31,Hemmert:1997ye} and list them in the second column of Table \ref{blbt}. The covariant derivatives acting on the meson fields and the external sources are counted as ${\mathcal O}(p^1)$, but those acting on the nucleon and $\Delta$ fields are counted as ${\mathcal O}(p^0)$. According to the low-energy approximation for the bilinear coupling, $\pi\Delta\Delta$ or $\pi N\Delta$, one assigns the chiral dimensions for the elements of the Clifford algebra, the Pauli matrices, and the Levi-Civita tensors in the second column of Table \ref{cabt} \cite{pin4,pib31,Scherer:2012xha}.

\begin{table*}[!h]
\caption{\label{blbt}Chiral dimension (Dim), parity ($P$), charge conjugation ($C$), and Hermiticity (h.c.) of the building blocks, where $c^{11}=-c^{22}=c^{33}=1$ and $c^{ij}=0$ when $i\neq j$. The building blocks in the first four rows are used in the chiral Lagrangian without $\Delta$. The building blocks in the last six rows are used in the chiral Lagrangian with $\Delta$.}
{\renewcommand\arraystretch{1.2}
\begin{tabular}{ccccc}
	\hline\hline
	                     & Dim &          $P$           &              $C$               &         h.c.          \\
	\hline
	     $u^{\mu}$       &  1  &       $-u_{\mu}$       &         $(u^{\mu})^T$          &       $u^{\mu}$       \\
	    $h^{\mu\nu}$     &  2  &     $-h_{\mu\nu}$      &        $(h^{\mu\nu})^T$        &     $h^{\mu\nu}$      \\
	    $\chi_{\pm}$     &  2  &    $\pm\chi_{\pm}$     &        $(\chi_{\pm})^T$        &   $\pm \chi_{\pm}$    \\
	 $f_{\pm}^{\mu\nu}$  &  2  &  $\pm f_{\pm\mu\nu}$   &   $\mp (f_{\pm}^{\mu\nu})^T$   &  $ f_{\pm}^{\mu\nu}$  \\
	    $u_i^{\mu}$      &  1  &      $-u_{i,\mu}$      &       $c_{ij}u_j^{\mu}$        &      $u_i^{\mu}$      \\
	   $h_i^{\mu\nu}$    &  2  &    $-h_{i,\mu\nu}$     &      $c_{ij}h_j^{\mu\nu}$      &    $h_i^{\mu\nu}$     \\
	   $\chi_{\pm,i}$    &  2  &   $\pm\chi_{\pm,i}$    &      $c_{ij}\chi_{\pm,j}$      &  $\pm \chi_{\pm,i}$   \\
	   $\chi_{\pm,s}$    &  2  &   $\pm\chi_{\pm,s}$    &         $\chi_{\pm,s}$         &  $\pm \chi_{\pm,s}$   \\
	$f_{\pm,i}^{\mu\nu}$ &  2  & $\pm f_{\pm,i\mu\nu}$ & $\mp c_{ij}f_{\pm,j}^{\mu\nu}$ & $ f_{\pm,i}^{\mu\nu}$ \\
	 $f_{+,s}^{\mu\nu}$  &  2  &    $ f_{+,s\mu\nu}$    &      $- f_{+,s}^{\mu\nu}$      &  $ f_{+,s}^{\mu\nu}$  \\
	\hline\hline
\end{tabular}
}
\end{table*}

\begin{table*}[!h]
\caption{\label{cabt}Chiral dimension (Dim), parity ($P$), charge conjugation ($C$), and Hermiticity (h.c.) of the Clifford algebra elements, the Levi-Civita tensors, and the Pauli matrices. The subscript `$\Delta\Delta$' (`${N\Delta}$') denotes the $\pi\Delta\Delta$ ($\pi N\Delta$) case. The meaning of the plus or minus sign is explained in the text.}
\begin{tabular}{cccccccc}
	\hline\hline
	                                  & Dim & $P_{\Delta\Delta}$ & $C_{\Delta\Delta}$ & h.c.$_{\Delta\Delta}$ & $P_{N\Delta}$ & $C_{N\Delta}$ & h.c.$_{N\Delta}$ \\ \hline
	               $1$                &  0  &    $+$     &    $+$     &      $+$      &      $-$      &      $+$      &       $+$        \\
	             $\gamf$              &  1  &    $-$     &    $+$     &      $-$      &      $+$      &      $+$      &       $-$        \\
	         $\gamma^{\mu}$           &  0  &    $+$     &    $-$     &      $+$      &      $-$      &      $-$      &       $+$        \\
	         $\fgamma^{\mu}$          &  0  &    $-$     &    $+$     &      $+$      &      $+$      &      $+$      &       $+$        \\
	        $\sigma^{\mu\nu}$         &  0  &    $+$     &    $-$     &      $+$      &      $-$      &      $-$      &       $+$        \\
	$\varepsilon^{\mu\nu\lambda\rho}$ &  0  &    $-$     &    $+$     &      $+$      &      $-$      &      $+$      &       $+$        \\
	        $\epsilon^{ijk}$          &  0  &    $+$     &    $-$     &      $+$      &      $+$      &      $-$      &       $+$        \\
	            $\tau_i$              &  0  &    $+$     &    $+$     &      $+$      &      $+$      &      $+$      &       $+$        \\
	     $D^\mu \psi_{i}^\nu$      &  0  &    $+$     &    $-$     &      $-$      &      $+$      &      $+$      &       $+$        \\ \hline\hline
\end{tabular}
\end{table*}

The chiral Lagrangian needs to be invariant under the chiral rotation (R), parity transformation ($P$), charge conjugation transformation ($C$), and Hermitian transformation (h.c.). It is necessary to know these transformation properties for the building blocks and other essential elements.

Under the chiral rotation $R$, the transformation for the nucleon doublet is
\begin{eqnarray}\label{chiraltransN}
\psi&\xrightarrow{R}&\psi'=h\psi
\end{eqnarray}
and those for the RS fields are \cite{Tang:1996sq}
\begin{eqnarray}\label{chiralpsimu}
\psi_i^{\mu}&\xrightarrow{R}&K_{ij}h\psi_j^{\mu} \qquad (i,j=1,2,3)
\end{eqnarray}
where $K_{ij}=\frac{1}{2}\la\tau_i h\tau_jh^\dag\ra$. $D^\mu\psi_i^{\nu}$ also transforms in the same way. From the definitions in Eqs. \eqref{df} and \eqref{cd}, we have the following chiral transformations for the building blocks (and with their covariant derivatives)
\begin{eqnarray}
&X\xrightarrow{R} X'=hXh^\dag.&
\end{eqnarray}
With the decomposition in Eq. \eqref{Xi}, one gets the transformation properties for $X_i$ and $X_s$ as follows,
\begin{eqnarray}\label{chiralx}
&X^i\xrightarrow{R} K^{ij} X_j,\qquad X_s\xrightarrow{R}X_s.&
\end{eqnarray}
From Eqs. \eqref{chiraltransN}, \eqref{chiralpsimu}, and \eqref{chiralx} and the property $K^\dag=K^{-1}$, it is obvious that a structure like $\bar{\psi}X^iD_\mu\psi^\mu_i $ is chirally invariant, where the isovector indices are contracted.

Now we move on to the parity, charge conjugation, and Hermitian transformations. The transformation properties of the building blocks $X$'s are simple \cite{p62,pin4,pib31} and we collect them in the first four rows of Table \ref{blbt}. Because the parities of the Pauli matrices are even and $X=X_i\tau_i+X_sI_2$, $X_i$ and $X_s$ have the same parities as $X$. To consider the charge conjugation transformations, we adopt the properties for the Pauli matrices used in Ref. \cite{Hemmert:1997wz}:
$\tau_i^T=c^{ij}\tau_j$, where $c^{11}=-c^{22}=c^{33}=1$ and $c^{ij}=0$ when $i\neq j$. If the charge conjugation transformation of $X$ is $X\xrightarrow{C} (-1)^cX^T$, one has
\begin{eqnarray}
X^i\xrightarrow{C} (-1)^c c^{ij}X_j,\qquad X_s\xrightarrow{C}(-1)^c X_s.
\end{eqnarray}
Since the Pauli matrices are hermitian, the Hermitian transformations of $X_i$ and $X_s$ are the same as that of $X$. All of these transformation properties are collected in the last six rows of Table \ref{blbt}.

For the essential Clifford algebra elements and the Levi-Civita tensors in chiral Lagrangians, they are invariant under the chiral symmetry, but their parity, charge conjugation, and Hermitian transformation properties rely on the coupling structure, $\pi\Delta\Delta$ or $\pi N\Delta$. There are some differences between these two cases. Such transformations for the RS field are correlated with the Clifford algebra elements and also rely on the coupling structure. In the following, we discuss their transformation properties.

We here adopt the similar method used in Refs. \cite{pin4,Hemmert:1997ye} to analyze the transformations. In general, the invariant monomials of the $\pi\Delta\Delta$ chiral Lagrangian have the form
\begin{align}
\psib_i^\mu A^{ij}_{\cdots}\Theta_{\cdots}\psi_j^\nu+\mathrm{H.c.},\label{form1}
\end{align}
where ``$\cdots$'' denote some suitable Lorentz indices, $A^{ij}_{\cdots}$ is the product of $X_i$ and $X_s$, and $\Theta_{\cdots}$ is the product of a Clifford algebra element $\Gamma\in\{1,\gamma_\mu,\gamf,\fgamma_\mu,\sigma_{\mu\nu}\}$, the Levi-Civita tensors $\varepsilon^{\alpha\beta\rho\tau}$ and $\epsilon^{ijk}$, and several covariant derivatives $D_{\lambda}D_{\eta}\cdots$ acting on $\psi_j^\nu$. For the $\pi N\Delta$ chiral Lagrangian, the invariant monomials have the form
\begin{align}\label{form2}
\psib A^{i}_{\cdots}\Theta_{\cdots}\psi_{A,n,i}^\mu+\mathrm{H.c.}.
\end{align}
The meanings of the symbols are the same as those in the $\pi\Delta\Delta$ invariant monomials.

In Table \ref{cabt}, we list the parity, charge conjugation, and Hermitian transformation properties for the Clifford algebra elements, the Levi-Civita tensors, the Pauli matrices, and the covariant derivative of the RS field in these two cases. The positions of the subscript or superscript indices may be changed in the transformations and we present there only extra signs that one need to consider. Here follows some explanations.

Under the parity transformation, we have
\begin{eqnarray}
\psi\xrightarrow{P}\gamma_0\psi,\qquad \psi^i_{\mu}\xrightarrow{P} -\gamma_0\psi^{i,\mu},
\end{eqnarray}
where the extra minus sign comes from the explicit representation of the spin-$\frac{3}{2}$ fields given in Appendix A of Ref. \cite{Hemmert:1997ye}. For convenience, we absorb this minus sign into the Clifford algebra elements. Thus, a sign difference for the Clifford algebra elements exists between the $\pi\Delta\Delta$ case and the $\pi N\Delta$ case.

Under the charge conjugation transformation, the baryon fields transform as
\begin{align}
\psi\xrightarrow{C} -i(\bar\psi\gamma^0\gamma^2)^T,\quad \psi^{i,\mu}\xrightarrow{C} -ic^{ij}(\bar\psi^{j,\mu}\gamma^0\gamma^2)^T.\label{bc}
\end{align}
The factor $c^{ij}$ can be removed by $c^{ik}c^{kj}=\delta^{ij}$, which ensures the invariance of the Lagrangian. The properties of the Clifford algebra elements are the same as the $\pi NN$ case \cite{pin4}, but one should note the sign difference for $D^\mu \psi^\nu_i$. In our convention (Eqs. \eqref{form1} and \eqref{form2}), in the $\pi\Delta\Delta$ case, all covariant derivatives acting on $\psi_i^\mu$ will act on $\psi_i^\mu$ again after the $C$ transformation is imposed, while they, in the $\pi N\Delta$ case, still act on $\psib_{A,n,i}^\mu$ after the $C$ transformation. As a result, a sign difference between the $\pi\Delta\Delta$ and $\pi N\Delta$ cases exists (see Sec. \ref{pirl}), which is shown in the last row of Table \ref{cabt}. Of course, a different convention does not affect the final result of chiral Lagrangian. The remaining transformation is for the Levi-Civita symbol $\epsilon^{ijk}$ which is usually charge invariant. In Table \ref{cabt}, for convenience, an extra minus sign is added which comes from the determinant of $c_{ij}$,
\begin{eqnarray}
&\epsilon^{ijk}X_iY_jZ_k\xrightarrow{C}\epsilon^{ijk}(-1)^{x}c_{ii'}X_{i'}(-1)^{y}c_{jj'}Y_{j'}(-1)^{z}c_{kk'}Z_{k'}\notag\\
=&(-1)^{x+y+z}\det(c_{mn})\epsilon^{ijk}X_iY_jZ_k=-(-1)^{x+y+z}\epsilon^{ijk}X_iY_jZ_k,\label{fc}
\end{eqnarray}
where $x$, $y$, and $z$ are the $C$-parities of $X_i$, $Y_j$, and $Z_k$ in Table \ref{blbt}, respectively.

Under the Hermitian transformation, the signs for a Clifford algebra element in the $\pi\Delta\Delta$, $\pi N\Delta$, and $\pi NN$ cases \cite{pin4} are the same, while the sign difference for $D^\mu \psi^\nu_i$ is the same as $C$ transformations discussed above, which is shown in the last row of Table \ref{cabt}.

\subsection{Linear relations}\label{lr}
In constructing chiral Lagrangians, one needs to find out all independent monomials which are invariant under various transformations. Several linear relations are proved to be useful. We collect all independent linear relations as follows. Their hermitian relations will not be given explicitly.

\subsubsection{Subsidiary condition}

With the relations for the Pauli matrices and the Levi-Civita tensors, e.g.
\begin{align}
\begin{split}
&\tau_i\tau_j=\delta_{ij}+i\epsilon_{ijk}\tau^k,\\
&\tau_i\tau_j\tau_k=\delta_{jk}\tau_i+\delta_{ij}\tau_k-\delta_{ik}\tau_j+i\epsilon_{ijk},\\
&\epsilon^{ijk}\epsilon^{lmn}=\begin{vmatrix}
\delta^{il} & \delta^{im} & \delta^{in}\\
\delta^{jl} & \delta^{jm} & \delta^{jn}\\
\delta^{kl} & \delta^{km} & \delta^{kn}
\end{vmatrix},
\end{split}\label{paulir}
\end{align}
more complicated structures can be simplified to that with one Pauli matrix and one Levi-Civita tensor at most. Together with the subsidiary condition $\tau^i\psi^\mu_i=0$, terms like $\epsilon^{ijk}\psi^\mu_{k}$ and $\epsilon^{ijk}\tau_k\psi^\mu_{j}$ can be removed because
\begin{eqnarray}\label{epsl2}
\epsilon^{ijk}\psi^\mu_{k}&=&i\tau^i\psi^{j\mu}-i\tau^j\psi^{i\mu},\nonumber\\
\epsilon^{ijk}\tau_k\psi^\mu_{j}&=&(-i\tau^i\tau^j+i\delta^{ij})\psi^\mu_{j}=i\psi^{\mu}_{i}.
\end{eqnarray}
The former equation (the left-hand side without any Pauli matrix) means that a term containing $\epsilon^{ijk}$ whose index is contracted with the RS field is the linear combination of terms with one Pauli matrix. The latter equation means that a term containing $\epsilon^{ijk}$ whose indices are contracted with a Pauli matrix and the RS field is proportional to $\psi_i^{\mu}$. Therefore, the invariant monomials with contraction structures on the left hand sides can be removed. These relations reduce most contraction possibilities for monomials containing $\epsilon^{ijk}$. For a term with this Levi-Civita tensor, we only need to consider the case that the index contraction occurs among $\epsilon^{ijk}$, $X_i$'s, and one (at most) Pauli matrix.

Since the baryon field also contains the isospin index, more relations constraining the chiral Lagrangians are possible. Combine the two equations in Eq. \eqref{epsl2} and the last one in \eqref{paulir}, one can remove the Levi-Civita tensor and obtain, up to the $\mathcal{O}(p^4)$ order, the following relation.
\begin{align}
\psib_i^\mu\tau_jO^{ijk}_{\mu\nu}\psi_k^\nu+\psib_k^\mu\tau_jO^{ijk}_{\mu\nu}\psi_i^\nu-\psib_j^\mu\tau_iO^{ijk}_{\mu\nu}\psi_k^\nu-\psib_k^\mu\tau_iO^{ijk}_{\mu\nu}\psi_j^\nu
+\psib_l^\mu\tau_i\delta_{kj}O^{ijk}_{\mu\nu}\psi_l^\nu-\psib_l^\mu\tau_j\delta_{ki}O^{ijk}_{\mu\nu}\psi_l^\nu=0,
\end{align}
where $O^{ijk}_{\mu\nu}$ does not contain $\epsilon^{ijk}$ and the Pauli matrix.

\subsubsection{Schouten identity}

For the Levi-Civita tensors $\varepsilon^{\mu\nu\lambda\rho}$ and $\epsilon^{ijk}$ appearing in the invariant monomials, one has the Schouten identities,
\begin{eqnarray}
&\varepsilon^{\mu\nu\lambda\rho}A^\sigma-\varepsilon^{\sigma\nu\lambda\rho}A^\mu-\varepsilon^{\mu\sigma\lambda\rho}A^\nu
-\varepsilon^{\mu\nu\sigma\rho}A^\lambda-\varepsilon^{\mu\nu\lambda\sigma}A^\rho=0,&\nonumber\\
&\epsilon_{ijk}A_l-\epsilon_{ljk}A_i-\epsilon_{ilk}A_j-\epsilon_{ijl}A_k=0.&\label{si}
\end{eqnarray}
Combining the second identity with equations in \eqref{epsl2}, we obtain two more relations
\begin{align}
&\epsilon^{ijk}{\cal O}^l\psi_{l}^{\mu}=i{\cal O}_i\tau_j\psi^\mu_{k}+\mathrm{P}(i,j,k),\nonumber\\
&i{\cal O}_l\psi^{\mu}_{i}-i{\cal O}_i\psi^{\mu}_{l}-\epsilon^{ilk}{\cal O}_j\tau_k\psi^\mu_{j}-\epsilon^{ijl}{\cal O}_k\tau_k\psi^\mu_{j}=0,
\end{align}
where $P(i,j,k)$ means all permutations for the indices $i$, $j$, and $k$ and an odd permutation $P(i,j,k)$ gives a minus sign. The former equation indicates that we can remove terms with $\epsilon^{ijk}$ (there is no Pauli matrix in ${\cal O}^l$), while the latter equation (together with Eq. \eqref{epsl2}) indicates that one of the last two terms on the left-hand side can be removed. Several similar relations about Lorentz indices also exist, which is briefly discussed in the item \ref{extra1} of Sec. \ref{beom}.

\subsubsection{Partial integration}\label{pirl}

The partial derivative acting on the whole monomial, $\partial^\mu(monomial)$, can be discarded and one has
\begin{eqnarray}
0&=&\psib^{i\nu}\accentset{\leftharpoonup}{D}^\mu {\cal O}^{\cdots}\psi^{k\lambda}+\psib^{i\nu}\nabla^\mu {\cal O}^{\cdots}\psi^{k\lambda}+\psib^{i\nu} {\cal O}^{\cdots}D^\mu\psi^{k\lambda},\nonumber\\
0&=&\psib\accentset{\leftharpoonup}{D}^\mu {\cal O}^{\cdots}\psi^{j\nu}+\psib\nabla^\mu {\cal O}^{\cdots}\psi^{j\nu}+\psi {\cal O}^{\cdots}D^\mu\psi^{j\nu},
\end{eqnarray}
where ${\cal O}^{\cdots}$ is the product of $A_{\cdots}$ and $\Theta_{\cdots}$ in \eqref{form1} or \eqref{form2} and ``$\cdots$'' represent suitable indices. Because the covariant derivative acting on the nucleon and $\Delta$ fields is counted as ${\mathcal O}(p^0)$ and that on $X_i$ or $X_s$ is ${\mathcal O}(p^1)$, we can simply employ the following relations in reducing the number of monomials \cite{pin4,pib31},
\begin{eqnarray}\label{partialintegration}
&\psib^{i\nu}\accentset{\leftharpoonup}{D}^\mu {\cal O}^{\cdots}\psi^{k\lambda}\doteq-\psib^{i\nu} {\cal O}^{\cdots}D^\mu\psi^{k\lambda},&\nonumber\\
&\psib\accentset{\leftharpoonup}{D}^\mu {\cal O}^{\cdots}\psi^{j\nu}\doteq-\psi {\cal O}^{\cdots}D^\mu\psi^{j\nu}.&
\end{eqnarray}
The symbol ``$\doteq$" means that both sides are equal if high order terms are ignored. For the purpose of constructing Lagrangian in a unified way, we choose a convention for the position of the covariant derivative acting on the RS field. In the $\pi\Delta\Delta$ case, we move all covariant derivatives to the right-side $\Delta$ field. In the $\pi N\Delta$ case, we move all the covariant derivatives to the $\Delta$ field no matter whether it is on the left side or on the right side of a monomial. This convention results in the sign difference for the charge conjugation and Hermitian transformations of $D^\mu\psi_i^\nu$ between these two cases, which has been shown in Table \ref{cabt}.

\subsubsection{Equations of motion (EOM)}\label{beom}
The lowest order EOM from the pseudoscalar chiral Lagrangian is
\begin{align}
\nabla_\mu u^\mu&=\frac{i}{2}\bigg(\chim-\frac{1}{N_f}\la\chim\ra\bigg),\label{eomb}
\end{align}
where $N_f$ is the number of quark flavors and we take $N_f=2$ here. This equation indicates that the monomials including $\nabla_\mu u^\mu$ can be eliminated. Obviously, the higher order EOM has additional terms on the right hand side and they have no effects on the construction of chiral Lagrangian. In Ref. \cite{pin4}, the EOM from the $\pi NN$ chiral Lagrangian has been used to restrict the structures of $\Theta_{\cdots}$ to a small set (see also Ref. \cite{Jiang:2016vax}). Here, we constrain $\Theta_{\cdots}$ in Eq. \eqref{form1} or \eqref{form2} in a similar way.

When the interactions of $\Delta$ exist, the general $\pi\Delta\Delta$ chiral Lagrangian has the form \cite{Hemmert:1997ye}
\begin{eqnarray}
\mathscr{L}_{\pi\Delta}&=&\bar{\psi}^i_{\mu}\Lambda_{A,ij}^{\mu\nu}\psi^j_\nu,\\
\Lambda_{A,ij}^{\mu\nu}&=&-\big[(i\slashed{D}-m_\Delta)g^{\mu\nu}+iA(\gamma^\mu D^\nu+\gamma^\nu D^\mu)\nonumber\\
&&+\frac{i}{2}(3A^2+2A+1)\gamma^\mu\slashed{D}\gamma^\nu\nonumber\\
&&+m_\Delta(3A^2+3A+1)\gamma^\mu\gamma^\nu\big]\delta_{ij}+O_{1,ij}^{\mu\nu},\label{gcl}
\end{eqnarray}
where $O_{1,ij}^{\mu\nu}$ denotes terms containing pion fields and external sources. Each term in $O_{1,ij}^{\mu\nu}$ contains at least one building block in Table \ref{blbt}. Hence, $O_{1,ij}^{\mu\nu}$ is at least at the order $\mathcal{O}(p^1)$. With the same technique to obtain Eqs. \eqref{eompsif}-\eqref{eoms2}, a similar EOM and two subsidiary conditions are obtained,
\begin{align}
&(i\slashed{D}-m_\Delta)\psi_{i}^\mu\doteq 0,\label{eomdi}\\
&D_\mu\psi_{i}^\mu\doteq 0,\label{eomdis1}\\
&\gamma_\mu\psi_{i}^\mu\doteq 0.\label{eomdis2}
\end{align}
The strict forms on the right-hand sides of the above equations come from $O_{1,ij}^{\mu\nu}$ and they are at least at the order $\mathcal{O}(p^1)$. This even works if the discussion is only in the $\mathcal{O}(p^1)$ order because we only use building-block-independent terms (the terms in the square bracket in Eq. \eqref{gcl}) to obtain the above equations.

If we replace $\psi_i^\mu$ (isospin doublet) with $\psi$ (isospin doublet) in \eqref{eomdi}, one gets the nucleon EOM in Ref. \cite{pin4}. This correspondence indicates that all the derivation techniques used in Ref. \cite{pin4} can be applied to the present case. Hence, we may borrow directly the results obtained there. In addition to these $\pi NN$-like structures of $\Theta_{\cdots}$, the existence of the Lorentz index in the RS field results in more possibilities. Fortunately, the vector-spinor representation of spin-3/2 fields has two subsidiary conditions and they may be used to remove some structures. If we multiply $\Gamma$ $(\Gamma\in\{1,\gamma_\mu,\gamf,\fgamma_\mu,\sigma_{\mu\nu}\})$ in both sides of Eq. \eqref{eomdis2}, we obtain
\begin{align}
0\doteq \Gamma\gamma_\mu\psi_{i}^\mu=\sum_a \Gamma_{a,\mu}\psi_{i}^\mu,
\end{align}
where $\Gamma_{a,\mu}$ denotes the elements in $\{1,\gamma_\mu,\gamf,\fgamma_\mu,\sigma_{\mu\nu}\}$ ($g^{\mu\nu}$ and $\varepsilon^{\mu\nu\lambda\rho}$ may also be a part of $\Gamma_{a,\mu}$). This equation gives some similar relations as those from \eqref{eomdi} discussed above, which ensures that the Lorentz index of $\psi_i^\mu$ can be treated as the index of a covariant derivative acting on the nucleon in the $\pi NN$ case, i.e. the correspondence $\psi^\mu_i\leftrightarrow D^\mu\psi$ may be adopted (except Eq. \eqref{Dsys} below). Because of Item \ref{itemgamma} below, the relations coming from Eq. \eqref{eomdis1} are the same as those from Eq. \eqref{eomdis2}. By using the baryon EOMs and the subsidiary conditions, one gets all constraint conditions in constructing chiral Lagrangians. We summarize the constraint conditions as follows.
\begin{enumerate}
\item \label{itemgamma}The terms containing $\gamma^\mu$ can be changed to those with one more covariant derivative and the structure $\gamma^\mu$ alone does not appear in the Lagrangian.

\item The Lorentz indices of $D$'s are different from that of $\psi_{i}^\mu$ or $\bar{\psi}_{i}^\mu$.

\item \label{symd} The indices of $D$'s are totally different and totally symmetric. To reflect the symmetric nature, we use the short notation $D_{\nu\lambda\rho\cdots}$ to denote multiple derivatives where
\begin{align}
D_{\nu\lambda\rho\cdots}=D_\nu D_\lambda D_\rho\cdots+ \text{full permutation of $D$'s.} \label{Dsys}
\end{align}

\item When $\epsilon^{\mu\nu\lambda\rho}$ exists, neither $\fgamma^{\mu}$ nor $\sigma^{\mu\nu}$ exists because the combination can be converted to  structures like $(\sigma^{\mu\nu}D^\rho+\cdots)$ or $(\gamma_5\gamma^\mu D^\nu+\cdots)$.

\item \label{gamind}The Lorentz indices of $\fgamma^\mu$ and $\sigma^{\mu\nu}$ are different from that of $\psi_{i}^\mu$ or $\bar{\psi}_{i}^\mu$ and that of covariant derivative acting on the baryon fields.

\item \label{gamd}If $A^{\mu\nu\cdots}=\nabla^\mu B^{\nu\cdots}$ in Eq. \eqref{form1} or \eqref{form2}, the index of the covariant derivative $\nabla^\mu$ is different from that of the RS field or that of $D^\nu$ acting on the $\Delta$ or nucleon fields. The contraction structure that the index of $\nabla^\mu$ is the same as that of $\psi_\mu$ or $D_\mu$ vanishes or has equivalent descriptions if high order terms are ignored.

\item \label{extra1} In constructing high order chiral Lagrangians, more relations coming from the Schouten identity are needed, see Eqs. (A7)-(A10) in Ref. \cite{Jiang:2016vax}. Simply speaking, the sum of all permutations of five different indices (one or two indices come from $\fgamma^\mu$ or $\sigma^{\mu\nu}$) vanishes up to high order terms, where an odd permutation gives a minus sign.

\item \label{itemvii} The difference between $\psi_{i}^\mu$ and $\psi_{Ai}^\mu$ is of higher order terms containing external sources. Hence, we could use $\psi_{Ai}^\mu$ instead of $\psi_{i}^\mu$ to construct the Lagrangians.

\item \label{itemviii} Because $\psi^\mu_i$ contains a Lorentz index, additional relations may be obtained. With Eq. \eqref{eomdis2} and the formula
\begin{align}
\gamma^\mu\gamma^\nu\gamma^\lambda=g^{\mu\nu}\gamma^\lambda
+g^{\nu\lambda}\gamma^\mu-g^{\lambda\mu}\gamma^\nu-i\varepsilon^{\mu\nu\lambda\rho}\gamma_{\rho}\gamma^5,
\end{align}
we get
\begin{align}
&m\varepsilon^{\mu\nu\lambda\rho}D_{\rho}\psi_{\lambda,i}\doteq\gamma^\mu\gamf\psi^{\nu}_{i}-\gamma^\nu\gamf\psi^{\mu}_{i},\\
&\sigma^{\nu\lambda}\psi_i^\mu+\sigma^{\lambda\mu}\psi_i^\nu+\sigma^{\mu\nu}\psi_i^\lambda\doteq0.\label{sigpsi}
\end{align}
The first equation means that the indices of $\varepsilon^{\mu\nu\lambda\rho}$ can not be contracted with that of a covariant derivative and that of the $\Delta$ field simultaneously. The second one means that one of the three terms on the left-hand side can be removed.
\end{enumerate}

Up to the $\mathcal{O}(p^4)$ order, all possible Lorentz structures of $\Theta_{\cdots}$ are constrained to be
\begin{align*}
\mathcal{O}(p^1)_{\pi N\Delta}:& 1,\\
\mathcal{O}(p^1)_{\pi \Delta\Delta},\mathcal{O}(p^2)_{\pi N\Delta}:& D^\mu,\fgamma^{\mu},\\
\mathcal{O}(p^2)_{\pi \Delta\Delta},\mathcal{O}(p^3)_{\pi N\Delta}:&1, D^{\mu\nu},\fgamma^{\mu}D^{\nu},\sigma^{\mu\nu},\epsilon^{\mu\nu\lambda\rho},\\
\mathcal{O}(p^3)_{\pi \Delta\Delta},\mathcal{O}(p^4)_{\pi N\Delta}:&D^\mu,D^{\mu\nu\lambda},\fgamma^{\mu},\fgamma^{\mu}D^{\nu\lambda},\sigma^{\mu\nu}D^{\lambda},\epsilon^{\mu\nu\lambda\rho}D_{\rho},\epsilon^{\mu\nu\lambda\rho}D^{\sigma},\\
\mathcal{O}(p^4)_{\pi \Delta\Delta}\phantom{,\mathcal{O}(p^4)_{\pi N\Delta}}:&1,D^{\mu\nu},D^{\mu\nu\lambda\rho}, \fgamma^{\mu}D^{\nu},\fgamma^{\mu}D^{\nu\lambda\rho},\sigma^{\mu\nu},\sigma^{\mu\nu}D^{\lambda\rho},\epsilon^{\mu\nu\lambda\rho},{\epsilon^{\mu\nu\lambda\rho}D_{\rho}}^{\sigma},\epsilon^{\mu\nu\lambda\rho}D^{\sigma\delta}.
\end{align*}

\subsubsection{Covariant derivatives and Bianchi identity}
From the relations in Eq. \eqref{Gam0} and \eqref{Gam1}, one gets
\begin{eqnarray}\label{bi}
\nabla^\mu\Gamma^{\nu\lambda}+\nabla^\nu\Gamma^{\lambda\mu}+\nabla^\lambda\Gamma^{\mu\nu}=0.
\end{eqnarray}
This Bianchi identity gives a relation between the covariant derivatives of $\Gamma^{\mu\nu}$ or $\fp^{\mu\nu}$. To reveal the isovector indices explicitly, we decompose Eqs. \eqref{Gam0} and \eqref{bi} with $\Gamma^{\mu\nu}=\Gamma^{\mu\nu}_i\tau_i+\Gamma^{\mu\nu}_sI_2$ and get
\begin{eqnarray}
&[\nabla^\mu,\nabla^{\nu}]X^i=-2i\epsilon^{ijk}X_j\Gamma_k^{\mu\nu},&\nonumber\\
&\nabla^\mu\Gamma_i^{\nu\lambda}+\nabla^\nu\Gamma_i^{\lambda\mu}+\nabla^\lambda\Gamma_i^{\mu\nu}=0,&\nonumber\\
&\nabla^\mu\Gamma_{s}^{\nu\lambda}+\nabla^\nu\Gamma_{s}^{\lambda\mu}+\nabla^\lambda\Gamma_{s}^{\mu\nu}=0.&
\end{eqnarray}
The first equation means that the exchange of two covariant derivatives acting on a building block is related to $\Gamma_k^{\mu\nu}$. Because the right-hand side structure has been considered in constructing Lagrangians, one of the two covariant derivatives on the left hand side can be removed. In other words, we can treat the covariant derivatives acting on $X_i$ as commutative operators. The last two equations indicate that one of the three terms on the left hand side can be removed.

\subsubsection{Contact terms}
The contact terms involve only $\Delta$ and nucleon fields and pure external sources ($F_R^{\mu\nu}$, $F_L^{\mu\nu}$, $\chi$, and $\chi^\dag$). The number of such terms is small and we construct them separately. To adopt the above constraint relations, we also use the following formulas by revealing explicitly the sources in Eq. \eqref{df},
\begin{align}
\begin{split}
F_L^{\mu\nu}&=\frac{1}{2}u^\dag(f_+^{\mu\nu}+f_-^{\mu\nu})u,\\
F_R^{\mu\nu}&=\frac{1}{2}u(f_+^{\mu\nu}-f_-^{\mu\nu})u^\dag,\\
\chi&=\frac{1}{2}u(\chi_++\chi_-)u,\\
\chi^\dag&=\frac{1}{2}u^\dag(\chi_+-\chi_-)u^\dag.
\end{split}\label{lrct}
\end{align}
Up to the fourth chiral order, only the ${\mathcal O}(p^4)$ $\pi\Delta\Delta$ Lagrangian contains contact terms. The total number of the contact terms is six and we list them in the last items in Table \ref{p4pideltab}.

\subsubsection{More}

Because the isovector indices of the building blocks are given explicitly, the Cayley-Hamilton relation used in the construction of meson or $\pi NN$ chiral Lagrangians is ignored. It has been implied in the Pauli matrix relations. No more relations need to be considered in constructing chiral Lagrangians with $\Delta$.

\subsection{Reduction of the monomials}\label{redm}

It is helpful to create some rules to reduce conveniently the constructed monomials with the above relations. Before constructing the chiral Lagrangian, we use the following rules to express the monomials in a unified form.

\begin{enumerate}
\item The symbol $\varepsilon^{\mu\nu\lambda\rho}$ is moved to the far left and follows an $\epsilon^{ijk}$, if they exist. Behind these Levi-Civita tensors is $\psib$ or $\psib_{i}^\mu$. The field $\psi_{i}^\mu$ (with its covariant derivatives) is moved to the far right. Between $\psib$ (or $\psib_{i}^\mu$) and $\psi_{i}^\mu$, the building blocks $X_i$ and $X_s$, the Pauli matrices, and the $\gamma$ matrices are placed in order. Because $X_i$ and $X_s$ are C-numbers, their positions in the Lagrangian are actually arbitrary.

\item To set down the positions of $X_i$ and $X_s$, we assign a number to each of them or to its covariant derivative by ignoring its Lorentz and isovector indices temporarily. One only cares about the relative magnitudes of the numbers and their absolute values are not important. Table \ref{noo} shows an example. Each combination of the building blocks is mapped to a vector. In the combination of two $u$'s and one $h$, for example, we have three permutations $uuh\to(121,121,151)$, $huu\to(151,121,121)$, and $uhu\to(121,151,121)$. Of course, they are not different in describing physical processes. In the construction of chiral Lagrangian, we choose the combination with the smallest vector where the smaller number is placed as far left as possible, $uuh$ in this example.

\begin{table*}[h]
\caption{\label{noo}Numbering examples for building blocks (ignoring indices) and their indices. Only the relative magnitudes (but not the absolute values) of the numbers are meaningful.}
\renewcommand\arraystretch{1.2}
\begin{tabular}{cccccccccccccc}

	\hline\hline
	   Operator     & $\psib(N)$ & $\psib(\Delta)$  & $\psi(\Delta)$ &  $u$   & $\nabla u$ &   $h$    & $\Gamma$ & $f_{-}$ & $\cdots$ &  &  &  &  \\
	\hline
	    Number      &  7301   &    7501     &    7521    &  121   &    122     &   151    &   101    &   181   & $\cdots$ &  &  &  &  \\
	\hline\hline
	 Lorentz Index  &  $\mu$  &    $\nu$    & $\lambda$  & $\rho$ &  $\sigma$  & $\cdots$ &          &         &          &  \\
	\hline
	    Number      &    1    &      2      &     3      &   4    &     5      & $\cdots$ &          &         &          &  \\
	\hline\hline
	Isovector Index &   $i$   &     $j$     &    $k$     &  $l$   &    $m$     & $\cdots$ &          &         &  \\
	\hline
	    Number      &   16    &     17      &     18     &   19   &     20     & $\cdots$ &          &         &  \\
	\hline\hline
\end{tabular}
\end{table*}

\item For the Lorentz and isovector indices, the rules are the same as the building blocks. All indices are numbered, too. We also give an example in Table \ref{noo}. After the places of all the building blocks are fixed, each type of indices is also mapped to a vector. In the case of Lorentz indices, for example, we have $\psib u_i^\mu u_j^\nu  h^i_{\mu\nu}\gamma^{\lambda}\psi_{\lambda}^{j}\to (1,2,1,2,3,3)$, $\psib u_i^\mu u_j^\nu  h^i_{\nu\mu}\gamma^{\lambda}\psi_{\lambda}^{j}\to (1,2,2,1,3,3)$, $\psib u_i^\nu u_j^\mu  h^i_{\mu\nu}\gamma^{\lambda}\psi_{\lambda}^{j}\to (2,1,1,2,3,3)$, and so on. The possible sign problem in this mapping is also considered. In the construction of chiral Lagrangians, we choose the permutation with the smallest vector, $\psib u_i^\mu u_j^\nu  h^i_{\mu\nu}\gamma^{\lambda}\psi_{\lambda}^{j}\to (1,2,1,2,3,3)$. In the case of isovector indices, a similar choice procedure is employed. We have used the Einstein summation convention, the commutation relations of C-numbers, and the symmetric or antisymmetric relations for $f^{\mu\nu}_{\pm,i}$, $h_i^{\mu\nu}$, $\varepsilon^{\mu\nu\lambda\rho}$ and so on in the above rule creation process.
\end{enumerate}

We say that a monomial obeying the above rules has a {\it standard form}. With these rules, two monomials having the same standard form are equal, and vice versa. Besides the purpose of distinguishing monomials, the standard form is also convenient in programming. The final results are all presented in this form.

\subsection{Classifications and Substitutions}

Although it is not difficult to obtain all possible invariant monomials at a given order, the number of these monomials is too large if the order is high and it makes the further calculation complex. A more efficient method is to classify all the monomials according to the numbers of external sources, Levi-Civita tensors, and Pauli matrices. It means that we can treat first the category with four pseudoscalar sources without any Levi-Civita tensors or Pauli matrices, then the category with three pseudoscalar sources plus one vector current (or one covariant derivative, see Eq. \eqref{cd}) without any Pauli matrices or Levi-Civita tensors, and so on. The reliability of this classification is ensured by the linear relations in Sec. \ref{lr}. From these relations, it is observed that only monomials in the same category can be related. However, we need to deal with the contact terms separately.

To simplify the calculation, we usually make the following replacements,
\begin{align}
f_{+i}^{\mu\nu}\longleftrightarrow i\Gamma_i^{\mu\nu},\hspace{0.2cm}
f_{+s}^{\mu\nu}\longleftrightarrow i\Gamma_s^{\mu\nu},\label{rdb}
\end{align}
which are acceptable since our purpose is only to construct the general Lagrangians. The differences induced by these replacements can be compensated by other terms containing $u_i^\mu$ at the same chiral order.

\subsection{Independent linear relations and chiral Lagrangians}

With the above preparations and the systematic approach to construct meson and meson-baryon chiral Lagrangians in Refs. \cite{ourf,Jiang:2016vax}, we now construct chiral Lagrangnians with $\Delta$ as follows.

First of all, because some linear relations in Sec. \ref{lr} contain covariant derivatives, it is convenient for us to reveal manifestly the covariant derivatives in $h_i^{\mu\nu}$ and $f_{-i}^{\mu\nu}$ through
\begin{align}
h_i^{\mu\nu}&=\nabla^{\mu}u_{i}^{\nu}+\nabla^{\nu}u_{i}^{\mu}\label{hu},\\
f_{-i}^{\mu\nu}&=-\nabla^{\mu}u_{i}^{\nu}+\nabla^{\nu}u_{i}^{\mu}\label{fmu}.
\end{align}
We use $D_{i,j}$ to store all possible invariant monomials constructed with $\psib$, $\psib_{i}^\mu$, $\psi_{i}^{\mu}$, $\psib_{A,n,i}^{\mu}$, $\psi_{A,n,i}^{\mu}$, $u_i^\mu$, $\chi_{\pm,i}$, $\chi_{\pm,s}$, $h_i^{\mu\nu}$, $\Gamma_i^{\mu\nu}$, $\Gamma_s^{\mu\nu}$, $f_{-i}^{\mu\nu}$, and their derivative forms and use $E_{i,j}$ to store all possible monomials revealing the covariant derivatives (constructed with $\psib$, $\psib_{i}^\mu$, $\psi_{i}^{\mu}$, $\psib_{A,n,i}^{\mu}$, $\psi_{A,n,i}^{\mu}$, $u_i^\mu$, $\chi_{\pm,i}$, $\chi_{\pm,s}$,  $\Gamma_i^{\mu\nu}$, $\Gamma_s^{\mu\nu}$, and their derivative forms). Here, the index $i$ labels the categories and the index $j$ labels the monomials inside the category $i$. The linear relations between $D_{i,j}$ and $E_{i,j}$ are
\begin{align}
D_{i,j}=\sum_{k}A_{i,jk}E_{i,k},\label{lr0}
\end{align}
where the coefficient matrix $A_{i}$ for the category $i$ is easy to obtain with Eqs. \eqref{hu} and \eqref{fmu}.

Next, by applying the linear relations in Sec. \ref{lr}, we can find the constraint relations about $E_{i,k}$,
\begin{align}
\sum_{k}R_{i,jk}E_{i,k}=0,\label{rls}
\end{align}
where $R_{i}$ is the linear relation matrix for the category $i$. Usually, not all of these relations are independent. To extract the independent ones, we transform the matrix $R_{i}$ to the reduced row echelon form (row canonical form) $S_i$. The rank of $R_{i}$ or $S_{i}$ is equal to the number of independent linear relations and each nonzero row-vector of $S_{i}$ gives a linear relation. That is, the independent constraint equations read
\begin{align}
\sum_{k}S_{i,jk}E_{i,k}=0.
\end{align}
With these constrains, Eq. \eqref{lr0} can be revised to the form
\begin{eqnarray}
D_{i,j}=\sum_{k}A'_{i,jk}E_{i,k},\label{lr1}
\end{eqnarray}
where the matrix $A'_{i}$ is from the matrices $A_{i}$ and $S_{i}$ after all linear dependent constraints are removed.

Then, one extracts the independent terms. Now, the independent terms in $D_{i}$ are corresponding to the independent rows of $A'_{i}$ or the independent columns of $A^{\prime T}_{i}$. Similar to the processing of Eq. \eqref{rls}, one transforms the matrix $A^{\prime T}_{i}$ to the reduced row echelon form. Then the labels of the independent terms in $D_i$ and thus the final results can be extracted. The standard form defined in Sec. \ref{redm} ensures that all the linear relations have been used and all the independent monomials of $E_{i,k}$ are really independent.

After that, one constructs the contact terms. Because such terms connect monomials in different categories, we collect all the $D_{i,j}$ and $E_{i,k}$ in two big column vectors $D'_{j}$ and $E'_{k}$, respectively, and collect all $A'_{i,jk}$ in a big diagonal block matrix $A'_{jk}$. By repeating the same steps from Eq. \eqref{lr0} to Eq. \eqref{lr1}, one gets the independent terms containing contact terms. In fact, such terms can be constructed by hand since the number is small.

Lastly, according to the hermiticity, one adds an extra $i$ to some terms to ensure that the LECs are real. The Lagrangian with the original building blocks is also recovered with \eqref{rdb}.

\section{Results and discussions}\label{results}

With the steps given above, we obtain the minimal chiral Lagrangians with $\Delta$ up to the order ${\mathcal O}(p^4)$. As a cross check, we have confirmed the $\pi NN$ Lagrangians obtained in Ref. \cite{pin4}. Because the building blocks there are different from ours, the following relations and Eq. \eqref{paulir} are employed in this confirmation process,
\begin{eqnarray}
\la XY\ra&=&2X^iY_i+2X_sY_s,\nonumber\\
\la XYZ\ra&=&2i\epsilon^{ijk}X_iY_jZ_k+2X^iY_iZ_s\nonumber\\
&&+2X^iZ_iY_s+2Y^iZ_iX_s+2X_sY_sZ_s.
\end{eqnarray}

\subsection{$\mathcal{O}(p^1)$ order}

At the lowest chiral order, the obtained $\pi\Delta\Delta$ Lagrangian is
\begin{align}\label{p1pid}
\mathscr{L}^{(1)}_{\pi\Delta\Delta}&=-\bar{\psi}^i_{\mu}\big[(i\slashed{D}-m_\Delta)g^{\mu\nu}+iA(\gamma^\mu D^\nu+\gamma^\nu D^\mu)
+\frac{i}{2}(3A^2+2A+1)\gamma^\mu\slashed{D}\gamma^\nu\notag\\
&\phantom{=}~+m_\Delta(3A^2+3A+1)\gamma^\mu\gamma^\nu\big]\psi^i_\nu
+c^{(1)}_1{\psib}^{ i\mu}u^{ j\nu}\tau_{ j}\gamf\gamma_{\nu}\psi_{i\mu},
\end{align}
while the result from Ref. \cite{Hemmert:1997ye} is
\begin{align}
\mathscr{L}^{(1)}_{\pi\Delta\Delta}&=-\bar{\psi}^i_{\mu}\big[(i\slashed{D}-m_\Delta)g^{\mu\nu}+iA(\gamma^\mu D^\nu+\gamma^\nu D^\mu)
+\frac{i}{2}(3A^2+2A+1)\gamma^\mu\slashed{D}\gamma^\nu\notag\\
&\phantom{=}~+m_\Delta(3A^2+3A+1)\gamma^\mu\gamma^\nu
+\frac{g_1}{2}\slashed{u}\gamf
+\frac{g_2}{2}(\gamma^\mu u^{\nu}+u^{\mu}\gamma^\nu)\gamf
+\frac{g_3}{2}\gamma^\mu\slashed{u}\gamf\gamma^\nu\big]\psi^i_{\nu}.
\end{align}
The $g_1$ term and the $c^{(1)}_1$ term have the same structure. However, the former Lagrangian does not contain the $g_2$ and $g_3$ terms. The nonexistence of the $g_2$ and $g_3$ terms comes from Eqs. \eqref{eomdis1} and \eqref{eomdis2} or the item \ref{gamind} in Sec. \ref{beom}. Recall that these terms are involved only when the RS field is off the mass shell where the spin-1/2 components contribute \cite{Hemmert:1997ye}. In the RS representation, these components are unphysical and their contributions do not enter the $S$-matrix elements \cite{Krebs:2008zb,Krebs:2009bf}. In Refs. \cite{Krebs:2008zb,Krebs:2009bf}, it has been proved that the redundant off-shell parameters can be absorbed into redefinitions of LECs and these two terms are not necessary. The relations $g_2=Ag_1$ and $g_3=-\frac12(1+2A+3A^2)g_1$ derived in Ref. \cite{Wies:2006rv} also indicate that they are not independent. 
Now, with the $\Delta$ EOM and subsidiary conditions in Eqs. \eqref{eomdi}-\eqref{eomdis2}, we have eliminated the spin-1/2 contributions from the Lagrangian. Therefore, this elimination procedure gives the same feature that the $g_2$ and $g_3$ terms are not necessary.

For the lowest order $\pi N\Delta$ Lagrangian, our result is the same as that in Ref. \cite{Hemmert:1997ye},
\begin{align}
\mathscr{L}^{(1)}_{\pi N\Delta}&=g_{\pi N\Delta}\psib u_i^{\mu}\psi_{A,n,\mu}^i+\mathrm{H.c.}.
\end{align}

\subsection{$\mathcal{O}(p^2)$ order}
The obtained $\mathcal{O}(p^2)$ $\pi\Delta\Delta$ chiral Lagrangian is written as
\begin{align}
\mathscr{L}^{(2)}_{\pi\Delta}=\sum_{n=1}^{11}c^{(2)}_n O^{(2)}_n.
\end{align}
There are 11 independent terms and we list them in Table \ref{p2pidel}. The number of the terms are exactly the same as that from Ref. \cite{Hemmert:1997ye}. Note that the original Lagrangian in Ref. \cite{Hemmert:1997ye} is given in the heavy baryon formalism. The relativistic form is (we have changed their notations to ours)
\begin{align}
\mathscr{L}^{(2)}_{\pi\Delta}=&a_1\psib_{Ai}^\mu\chi_{+,s}\psi^i_{A\mu}
-\frac{1}{2}a_2\psib_{Ai}^\mu u^{\nu}u^{\lambda}D_{\nu\lambda}\psi^i_{A\mu}
+a_3\psib_{Ai}^\mu u^{\nu}u_{\nu}\psi^i_{A\mu}
+\frac{i}{2}a_4\psib_{Ai}^\mu u^{\nu}u^{\lambda}\sigma_{\nu\lambda}\psi^i_{A\mu}\notag\\
&+a_5\psib_{Ai}^\mu(\chip-\chi_{+,s})\psi^i_{A\mu}
+a_6i\psib_{Ai}^\mu f_{+\mu\nu}\psi^{i\nu}_{A}
+\frac{1}{2}a_7i\psib_{Ai}^\mu f_{+s,\mu\nu}\psi^{i\nu}_{A}
+4a_8\psib_{Ai}^\mu u^i_\nu u_j^\nu \psi^{j}_{A\mu}\notag\\
&-2a_9\psib_{Ai}^\mu u^{i\nu} u^{j\lambda}D_{\nu\lambda}\psi_{Aj\mu}
+2a_{10}\psib_{Ai}^\mu (u^i_\mu u_j^\nu+u_{j\mu} u^{i\nu})\psi^{j}_{A\nu}
+2a_{11}\psib_{Ai}^\mu u^j_\mu u_j^\nu\psi^{i}_{A\nu}.
\end{align}
Their transition rules can be found from the following Eq. \eqref{hr}. In the third column of Table \ref{p2pidel}, we show the relations between the LECs based on the transition rules.

\begin{table}[!h]
\caption{\label{p2pidel}Terms in the $\mathcal{O}(p^2)$ $\pi\Delta\Delta$ chiral Lagrangian and the LEC relations between our $c^{(2)}_n$ and $a_n$ in Eq. (113) of Ref. \cite{Hemmert:1997ye}.}
\begin{tabular}{rcl}
\hline\hline
$n$ & $O^{(2)}_n$ & $c^{(2)}_n$ \\
\hline
1 & $\psib^{ i\mu}u_{ i\mu}u^{ j\nu}\psi_{ j\nu}$ & $2a_{10}+a_4$ \\
2 & $\psib^{ i\mu}{u_{ i}}^{\nu}{u^{ j}}_{\mu}\psi_{ j\nu}$ & $2a_{10}-a_4$ \\
3 & $\psib^{ i\mu}{u_{ i}}^{\nu}{u^{ j}}_{\nu}\psi_{ j\mu}$ & $4a_{8}$ \\
4 & $\psib^{ i\mu}{u^{ j}}_{\mu}{u_{ j}}^{\nu}\psi_{ i\nu}$ & $2a_{11}$ \\
5 & $\psib^{ i\mu}u^{ j\nu}u_{ j\nu}\psi_{ i\mu}$ & $a_{3}$ \\
6 & $\psib^{ i\mu}{u_{ i}}^{\nu}u^{ j\lambda}D_{\nu\lambda}\psi_{ j\mu}$ & $-4a_{9}$ \\
7 & $\psib^{ i\mu}u^{ j\nu}{u_{ j}}^{\lambda}D_{\nu\lambda}\psi_{ i\mu}$ & $-2a_{2}$ \\
8 & $i\psib^{ i\mu}{f_{s,+\mu}}^{\nu}\psi_{ i\nu}$ & $a_{6}+a_7/2$ \\
9 & $i\psib^{ i\mu}{{{f_{+}}^{ j}}_{\mu}}^{\nu}\tau_{ j}\psi_{ i\nu}$ & $a_{6}$ \\
10 & $\psib^{ i\mu}\chi_{+,s}\psi_{ i\mu}$ & $a_{1}$ \\
11 & $\psib^{ i\mu}{\chi_{+}}^{ j}\tau_{ j}\psi_{ i\mu}$ & $a_{5}$ \\
\hline\hline
\end{tabular}
\end{table}

For the $\mathcal{O}(p^2)$ $\pi N\Delta$ chiral Lagrangian, we obtain three independent terms,
\begin{align}
\mathscr{L}^{(2)}_{\pi N\Delta}=&d^{(2)}_1\psib u^{ i\mu}u^{ j\nu}\tau_{ i}\gamf\gamma_{\mu}\psi_{A,n, j\nu}
+d^{(2)}_2\psib u^{ i\mu}u^{ j\nu}\tau_{ i}\gamf\gamma_{\nu}\psi_{A,n, j\mu}
+d^{(2)}_3i\psib{f_{+}}^{ i\mu\nu}\gamf\gamma_{\mu}\psi_{A,n, i\nu}+\mathrm{H.c.}.
\end{align}
From Ref. \cite{Hemmert:1997ye}, a different set of relativistic terms is,
\begin{align}
\mathscr{L}^{(2)}_{\pi N\Delta}=&(-\frac{1}{2}b_1i\psib_{A,1,i}^{\mu}f_{+\mu\nu}^{i}\fgamma^{\nu}
+b_2i\psib_{A,2,i}^{\mu}f_{-\mu\nu}^{i}D^{\nu}+b_3i\psib_{A,3,i}^{\mu}\nabla_{\mu}u_{\nu}^{i}D^{\nu}\notag\\
&-\frac{1}{2}b_4\psib_{A,4,i}^{\mu}u^i_\mu u^\nu\fgamma_\nu
-\frac{1}{2}b_5\psib_{A,5,i}^{\mu}u_\mu u^{i\nu} \fgamma_\nu)\psi+\mathrm{H.c.}. \label{pindel2}
\end{align}
We find the following correspondence between these two sets of terms: $d_1^{(2)}\leftrightarrow b_4$, $d_2^{(2)}\leftrightarrow b_5$, and $d_3^{(2)}\leftrightarrow b_1$. Because of the item \ref{gamd} in Sec. \ref{beom}, the $b_2$ and $b_3$ terms can be eliminated. Ref. \cite{Long:2010kt} also points out that the $b_3$ term in Eq. \eqref{pindel2} is redundant.

\subsection{$\mathcal{O}(p^3)$ and $\mathcal{O}(p^4)$ orders}

We define the $\mathcal{O}(p^3)$ and $\mathcal{O}(p^4)$ chiral Lagrangians as
\begin{eqnarray}
\mathscr{L}^{(m)}_{\pi\Delta\Delta}&=&\sum_{n}c^{(m)}_n O^{(m)}_n,\label{pideln}\\
\mathscr{L}^{(m)}_{\pi N\Delta}&=&\sum_{n}d^{(m)}_n (P^{(m)}_n+\mathrm{H.c.}),\label{pindeln}
\end{eqnarray}
where $m=3$ or 4 denotes the chiral dimension, $c^{(m)}_n$ and $d^{(m)}_n$ are the LECs, and $O^{(m)}_n$ and $P^{(m)}_n$ are independent interaction terms. The results are listed in Appendix \ref{aeom}. In the $\mathcal{O}(p^3)$ ($\mathcal{O}(p^4)$) $\pi\Delta\Delta$ Lagrangian, there exist 38 (318) independent terms and we show them in Table \ref{p3pideltab} (\ref{p4pideltab}). The last six items in Table \ref{p4pideltab} are contact terms. In the $\mathcal{O}(p^3)$ ($\mathcal{O}(p^4)$) $\pi N\Delta$ Lagrangian, there are 33 (218) independent terms and we present them in Table \ref{p3pindeltab} (\ref{p4pindeltab}). Note that the $z_n$ parameters are different for the $\pi N\Delta$ Lagrangians at different orders, but we do not distinguish them explicitly in the former notations.

\subsection{Point transformation}

Now we move on to the point transformation (Eqs. \eqref{pt1} and \eqref{pt2}) for the constructed chiral Lagrangian. Under this transformation, the RS field is not invariant but the structure of the Lagrangian should be invariant. As a result, the LECs in the Lagrangian are dependent on the arbitrary unphysical parameter $A$ in the $\pi\Delta\Delta$ Lagrangian. To reduce the uncertainty, one can also adopt the redefined RS field $\psi_{Ai}^\mu\equiv O_A^{\mu\nu}\psi_{i\nu}=(g^{\mu\nu}+\frac12A\gamma^\mu\gamma^\nu)\psi_{i\nu}$ instead of $\psi_{i}^\mu$ as a building block in the construction of chiral Lagrangians. In this scheme, the RS field is point invariant and all the LECs are independent of $A$. The structure of the above $\pi\Delta\Delta$ chiral Lagrangian is not changed except the first part of Eq. \eqref{p1pid} which now turns into \cite{Pascalutsa:1994tp,Hemmert:1997ye}
\begin{align}
\mathscr{L}^{(1)}_{\pi\Delta\Delta}&=-\bar{\psi}^i_{A\mu}\big[(i\slashed{D}-m)g^{\mu\nu}-\frac{1}{4}\gamma^\mu\gamma^\lambda(i\slashed{D}-m)\gamma_\lambda\gamma^\nu\big]\psi^i_{A\nu}+\cdots.
\end{align}
The difference in the chiral Lagrangian caused by the replacement $\psi_{i}^\mu\to \psi_{Ai}^\mu$ can be ignored due to the item \ref{itemvii} in Sec. \ref{beom}. For the $\pi N\Delta$ chiral Lagrangian, we have presented the results with $\psi_{A,n,i}^{\mu}$. One should note that the definition $\psi_{A,n,i}^{\mu}=\Theta_{A,n}^{\mu\nu}(z_n)\psi_{i\nu}$ is adopted when we use the original RS field $\psi_{i}^\mu$ in the $\pi\Delta\Delta$ Lagrangian while the definition $\psi_{A,n,i}^{\mu}\equiv\Theta_{n}^{\mu\nu}(z_n)\psi_{Ai\nu}$ is adopted when we use the redefined RS field $\psi_{Ai}^\mu$. Here, $\Theta_{A,n}^{\mu\nu}(z_n)$ and $\Theta_{n}^{\mu\nu}(z_n)$ are defined in Eq. \eqref{z-para}.

\subsection{Heavy baryon projection}

In this subsection, we briefly discuss the heavy baryon formalism in the small scale expansion scheme and only give the correspondence between the leading structures in the nonrelativistic case and in the relativistic case. More details can be found in Refs. \cite{pin4,Hemmert:1997ye}.

To give a heavy baryon Lagrangian, the four-momentum $p_\mu$ of the baryon is written as
\begin{align}
p^\mu=m_0v^\mu+k^\mu,
\end{align}
where $m_0$ is the nucleon mass in the chiral limit, $v^\mu$ is the four-velocity with $v^2=1$, and $k^\mu$ is a small off-shell momentum. The nucleon field $\psi$ is projected into a `large' (light) component $N$ and a `small' (heavy) component $h$ while the RS field is projected into a `large' spin-3/2 component $T_i^\mu$, a `small' spin-3/2 component, and four spin-1/2 components. In the heavy baryon formalism, the mentioned two `small' components and the spin-1/2 components of the RS field are all integrated out and the chiral Lagrangian involves only the `large' components $N$ and $T_i^\mu$. This nonrelativistic reduction results in some changes for the $\Theta_{\cdots}$ in \eqref{form1} or \eqref{form2}. Between the relativistic and nonrelativistic interaction Lagrangians (the lowest kinetic term needs separate treatment), the rules of correspondence are
\begin{align}
\begin{split}
1&\longleftrightarrow 1,\\
\gamma^\mu &\longleftrightarrow v^\mu,\\
\fgamma^\mu&\longleftrightarrow -2S^\mu,\\
\sigma^{\mu\nu}&\longleftrightarrow 2\varepsilon^{\mu\nu\lambda\rho}v_\lambda S_\rho=-2i[S^\mu,S^\nu],\\
\psi&\longleftrightarrow N\\
\psi_i^\mu&\longleftrightarrow T_i^\mu,\\
D^\nu\psi_i^\mu&\longleftrightarrow -im_0v^\nu T_i^\mu,
\end{split}\label{hr}
\end{align}
where $S^\mu=\frac{i}{2}\gamf\sigma^{\mu\nu}v_\nu$ is the Pauli-Lubanski spin vector and $D^\mu$ is the covariant derivative acting on the RS field. With these rules, it is easy to obtain the chiral-invariant terms without recoil corrections in the heavy baryon formalism. To get the complete Lagrangians (including $1/m_0$ corrections) in this formalism, one needs the strict integration procedure given in Ref. \cite{Hemmert:1997ye}.

\section{Summary}\label{summ}

In this paper, we present a systematic method for the construction of chiral Lagrangians with $\Delta(1232)$. It is suitable for computer programming and has been applied to constructing meson \cite{ourf} and meson-baryon \cite{Jiang:2016vax} chiral Lagrangians. With this method, we complete the minimal chiral Lagrangians with $\Delta$, ${\cal L}_{\pi\Delta\Delta}$ and ${\cal L}_{\pi N\Delta}$, up to the ${\mathcal O}(p^4)$ (one-loop) order. We also briefly discuss the properties of the point transformation and the heavy baryon projection for the results.

With the chiral Lagrangians up to ${\cal O}(p^4)$ constructed in this paper, further studies in both the heavy baryon formalism and the relativistic formalism can be performed. Various studies in different processes are  needed to test whether ChPT with $\Delta$ works well or not.

\section*{Acknowledgments}

We thank Profs. Li-Sheng Geng and Bing-Wei Long for useful discussions. This work was supported by the National Science Foundation of China (NSFC) under Grants No. 11565004, No. 11775132, No. U1731239 and No. 11533003 , the National Basic Research Program (973Programme) of China 2014CB845800, the special funding for Guangxi distinguished professors (Bagui Yingcai and Bagui Xuezhe), the Foundation of Guangxi Key Laboratory for Relativistic Astrophysics and High Level Innovation Team and Outstanding Scholar Program in Guangxi Colleges.

\appendix
\section{Independent terms in $\mathcal{O}(p^3)$ and $\mathcal{O}(p^4)$ chiral Lagrangians with $\Delta(1232)$}\label{aeom}

\begin{longtable}{llllll}
\caption{Terms in the $\mathcal{O}(p^3)$ $\pi \Delta\Delta$ chiral Lagrangian where $O^{(3)}_n$ is defined in Eq. \eqref{pideln}.}\label{p3pideltab}\\
\hline\hline $n$ & $O^{(3)}_n$ & $n$ & $O^{(3)}_n$ & $n$ & $O^{(3)}_n$\\
\hline\endfirsthead

\hline\hline $n$ & $O^{(3)}_n$ & $n$ & $O^{(3)}_n$ & $n$ & $O^{(3)}_n$\\
\hline\endhead

\hline\hline 
\endfoot

\hline\endlastfoot

1 & $\psib^{ i\mu}u_{ i\mu}u^{ j\nu}u^{ k\lambda}\tau_{ j}\gamf\gamma_{\nu}\psi_{ k\lambda}$ & 14 & $\psib^{ i\mu}{u_{ i}}^{\nu}{{{f_{-}}^{ j}}_{\mu}}^{\lambda}D_{\lambda}\psi_{ j\nu}+\mathrm{H.c.}$ & 27 & $i\psib^{ i\mu}{{{f_{+}}^{ j}}_{\mu}}^{\nu}{u_{ j}}^{\lambda}\gamf\gamma_{\lambda}\psi_{ i\nu}$  \\
2 & $\psib^{ i\mu}u_{ i\mu}u^{ j\nu}u^{ k\lambda}\tau_{ j}\gamf\gamma_{\lambda}\psi_{ k\nu}+\mathrm{H.c.}$ & 15 & $\psib^{ i\mu}{u_{ i}}^{\nu}{{{f_{-}}^{ j}}_{\nu}}^{\lambda}D_{\lambda}\psi_{ j\mu}+\mathrm{H.c.}$ & 28 & $i\psib^{ i\mu}{f_{+ i}}^{\nu\lambda}u^{ j\rho}\gamf\gamma_{\nu}D_{\lambda\rho}\psi_{ j\mu}+\mathrm{H.c.}$  \\
3 & $\psib^{ i\mu}{u_{ i}}^{\nu}{u^{ j}}_{\mu}u^{ k\lambda}\tau_{ j}\gamf\gamma_{\lambda}\psi_{ k\nu}+\mathrm{H.c.}$ & 16 & $\psib^{ i\mu}{u^{ j}}_{\mu}{f_{- j}}^{\nu\lambda}D_{\nu}\psi_{ i\lambda}+\mathrm{H.c.}$ & 29 & $i\psib^{ i\mu}{f_{s,+\mu}}^{\nu}u^{ j\lambda}\tau_{ j}\gamf\gamma_{\nu}\psi_{ i\lambda}+\mathrm{H.c.}$  \\
4 & $\psib^{ i\mu}{u_{ i}}^{\nu}{u^{ j}}_{\mu}u^{ k\lambda}\tau_{ k}\gamf\gamma_{\lambda}\psi_{ j\nu}$ & 17 & $\psib^{ i\mu}u^{ j\nu}{f_{- j\mu}}^{\lambda}D_{\nu}\psi_{ i\lambda}$ & 30 & $i\psib^{ i\mu}{f_{s,+\mu}}^{\nu}u^{ j\lambda}\tau_{ j}\gamf\gamma_{\lambda}\psi_{ i\nu}$  \\
5 & $\psib^{ i\mu}{u_{ i}}^{\nu}{u^{ j}}_{\nu}u^{ k\lambda}\tau_{ j}\gamf\gamma_{\lambda}\psi_{ k\mu}+\mathrm{H.c.}$ & 18 & $\psib^{ i\mu}u_{ i\mu}h^{ j\nu\lambda}D_{\nu}\psi_{ j\lambda}+\mathrm{H.c.}$ & 31 & $i\psib^{ i\mu}\nabla^{\nu}{f_{s,+\nu}}^{\lambda}D_{\lambda}\psi_{ i\mu}$  \\
6 & $\psib^{ i\mu}{u_{ i}}^{\nu}{u^{ j}}_{\nu}u^{ k\lambda}\tau_{ k}\gamf\gamma_{\lambda}\psi_{ j\mu}$ & 19 & $\psib^{ i\mu}{u_{ i}}^{\nu}{{h^{ j}}_{\nu}}^{\lambda}D_{\lambda}\psi_{ j\mu}+\mathrm{H.c.}$ & 32 & $i\psib^{ i\mu}\nabla^{\nu}{{{f_{+}}^{ j}}_{\nu}}^{\lambda}\tau_{ j}D_{\lambda}\psi_{ i\mu}$  \\
\hline\hline
\end{longtable}

\begin{longtable}{llll}
\caption{\label{p4pideltab}Terms in the $\mathcal{O}(p^4)$ $\pi \Delta\Delta$ chiral Lagrangian where $O^{(4)}_n$ is defined in Eq. \eqref{pideln}.}\\
\hline\hline $n$ & $O^{(4)}_n$ & $n$ & $O^{(4)}_n$ \\
\hline\endfirsthead

\hline\hline $n$ & $O^{(4)}_n$ & $n$ & $O^{(4)}_n$ \\
\hline\endhead

\hline\hline 
\endfoot

\hline\endlastfoot

1 & $\psib^{ i\mu}u_{ i\mu}u^{ j\nu}u_{ j\nu}u^{ k\lambda}\psi_{ k\lambda}$ & 160 & $i\psib^{ i\mu}{f_{+ i}}^{\nu\lambda}{u^{ j}}_{\mu}u^{ k\rho}\tau_{ j}D_{\nu\rho}\psi_{ k\lambda}+\mathrm{H.c.}$  \\
2 & $\psib^{ i\mu}u_{ i\mu}u^{ j\nu}{u_{ j}}^{\lambda}{u^{ k}}_{\nu}\psi_{ k\lambda}+\mathrm{H.c.}$ & 161 & $i\psib^{ i\mu}{f_{+ i}}^{\nu\lambda}{u^{ j}}_{\mu}u^{ k\rho}\tau_{ k}D_{\nu\rho}\psi_{ j\lambda}+\mathrm{H.c.}$  \\
3 & $\psib^{ i\mu}{u_{ i}}^{\nu}{u^{ j}}_{\mu}{u_{ j}}^{\lambda}{u^{ k}}_{\nu}\psi_{ k\lambda}$ & 162 & $i\psib^{ i\mu}{f_{+ i}}^{\nu\lambda}{u^{ j}}_{\nu}u^{ k\rho}\tau_{ j}D_{\lambda\rho}\psi_{ k\mu}+\mathrm{H.c.}$  \\
4 & $\psib^{ i\mu}{u_{ i}}^{\nu}{u^{ j}}_{\mu}{u_{ j}}^{\lambda}{u^{ k}}_{\lambda}\psi_{ k\nu}+\mathrm{H.c.}$ & 163 & $i\psib^{ i\mu}{f_{+ i}}^{\nu\lambda}{u^{ j}}_{\nu}u^{ k\rho}\tau_{ k}D_{\lambda\rho}\psi_{ j\mu}+\mathrm{H.c.}$  \\
5 & $\psib^{ i\mu}{u_{ i}}^{\nu}{u^{ j}}_{\mu}u^{ k\lambda}u_{ k\lambda}\psi_{ j\nu}$ & 164 & $i\psib^{ i\mu}{{{f_{+}}^{ j}}_{\mu}}^{\nu}{u_{ i}}^{\lambda}u^{ k\rho}\tau_{ j}D_{\nu\lambda}\psi_{ k\rho}+\mathrm{H.c.}$  \\
6 & $\psib^{ i\mu}{u_{ i}}^{\nu}{u^{ j}}_{\nu}{u_{ j}}^{\lambda}{u^{ k}}_{\lambda}\psi_{ k\mu}$ & 165 & $i\psib^{ i\mu}{{{f_{+}}^{ j}}_{\mu}}^{\nu}{u_{ i}}^{\lambda}u^{ k\rho}\tau_{ j}D_{\nu\rho}\psi_{ k\lambda}+\mathrm{H.c.}$  \\
7 & $\psib^{ i\mu}{u_{ i}}^{\nu}{u^{ j}}_{\nu}u^{ k\lambda}u_{ k\lambda}\psi_{ j\mu}$ & 166 & $i\psib^{ i\mu}{{{f_{+}}^{ j}}_{\mu}}^{\nu}{u_{ i}}^{\lambda}u^{ k\rho}\tau_{ j}D_{\lambda\rho}\psi_{ k\nu}$  \\
8 & $\psib^{ i\mu}{u^{ j}}_{\mu}{u_{ j}}^{\nu}{u^{ k}}_{\nu}{u_{ k}}^{\lambda}\psi_{ i\lambda}$ & 167 & $i\psib^{ i\mu}{{{f_{+}}^{ j}}_{\mu}}^{\nu}{u_{ j}}^{\lambda}u^{ k\rho}\tau_{ k}D_{\nu\lambda}\psi_{ i\rho}+\mathrm{H.c.}$  \\
\hline\hline
\end{longtable}

\begin{longtable}{llllll}
\caption{\label{p3pindeltab}Terms in the $\mathcal{O}(p^3)$ $\pi N\Delta$ chiral Lagrangian where $P^{(3)}_n$ is defined in Eq. \eqref{pindeln}.}\\
\hline\hline $n$ & $P^{(3)}_n$ & $n$ & $P^{(3)}_n$ & $n$ & $P^{(3)}_n$\\
\hline\endfirsthead

\hline\hline $n$ & $P^{(3)}_n$ & $n$ & $P^{(3)}_n$ & $n$ & $P^{(3)}_n$\\
\hline\endhead

\hline\hline 
\endfoot

\hline\endlastfoot

1 & $\psib u^{ i\mu}u_{ i\mu}u^{ j\nu}\psi_{A,n, j\nu}$ & 12 & $\psib u^{ i\mu}{f_{-}}^{ j\nu\lambda}\tau_{ j}\gamf\gamma_{\mu}D_{\nu}\psi_{A,n, i\lambda}$ & 23 & $\psib{f_{s,+}}^{\mu\nu}u^{ i\lambda}\sigma_{\mu\nu}\psi_{A,n, i\lambda}$  \\
2 & $\psib u^{ i\mu}{u_{ i}}^{\nu}{u^{ j}}_{\mu}\psi_{A,n, j\nu}$ & 13 & $\psib u^{ i\mu}{f_{-}}^{ j\nu\lambda}\tau_{ j}\gamf\gamma_{\nu}D_{\mu}\psi_{A,n, i\lambda}$ & 24 & $i\psib{f_{+}}^{ i\mu\nu}{u^{ j}}_{\mu}\tau_{ i}\psi_{A,n, j\nu}$  \\
3 & $\psib u^{ i\mu}{u_{ i}}^{\nu}u^{ j\lambda}D_{\mu\nu}\psi_{A,n, j\lambda}$ & 14 & $\psib u^{ i\mu}{f_{-}}^{ j\nu\lambda}\tau_{ j}\gamf\gamma_{\nu}D_{\lambda}\psi_{A,n, i\mu}$ & 25 & $i\psib{f_{+}}^{ i\mu\nu}{u^{ j}}_{\mu}\tau_{ j}\psi_{A,n, i\nu}$  \\
4 & $\psib u^{ i\mu}{u_{ i}}^{\nu}u^{ j\lambda}D_{\mu\lambda}\psi_{A,n, j\nu}$ & 15 & $\psib u^{ i\mu}h^{ j\nu\lambda}\tau_{ i}\gamf\gamma_{\mu}D_{\nu}\psi_{A,n, j\lambda}$ & 26 & $i\psib{f_{+}}^{ i\mu\nu}u^{ j\lambda}\tau_{ i}D_{\mu\lambda}\psi_{A,n, j\nu}$  \\
5 & $i\psib u^{ i\mu}{u_{ i}}^{\nu}u^{ j\lambda}\sigma_{\mu\lambda}\psi_{A,n, j\nu}$ & 16 & $\psib u^{ i\mu}h^{ j\nu\lambda}\tau_{ i}\gamf\gamma_{\nu}D_{\mu}\psi_{A,n, j\lambda}$ & 27 & $i\psib{f_{+}}^{ i\mu\nu}u^{ j\lambda}\tau_{ j}D_{\mu\lambda}\psi_{A,n, i\nu}$  \\
6 & $i\epsilon^{ i j k}\psib{u_{ i}}^{\mu}{u_{ j}}^{\nu}{u^{ l}}_{\mu}\tau_{ k}\psi_{A,n, l\nu}$ & 17 & $\varepsilon^{\mu\nu\lambda\rho}\psib{u^{ i}}_{\mu}{{f_{-}}^{ j}}_{\nu\lambda}\tau_{ i}\psi_{A,n, j\rho}$ & 28 & $\psib{f_{+}}^{ i\mu\nu}u^{ j\lambda}\tau_{ i}\sigma_{\mu\nu}\psi_{A,n, j\lambda}$  \\
7 & $i\epsilon^{ i j k}\psib{u_{ i}}^{\mu}{u_{ j}}^{\nu}u^{ l\lambda}\tau_{ k}D_{\mu\lambda}\psi_{A,n, l\nu}$ & 18 & $\varepsilon^{\mu\nu\lambda\rho}\psib{u^{ i}}_{\mu}{{f_{-}}^{ j}}_{\nu\lambda}\tau_{ j}\psi_{A,n, i\rho}$ & 29 & $\psib{f_{+}}^{ i\mu\nu}u^{ j\lambda}\tau_{ j}\sigma_{\mu\nu}\psi_{A,n, i\lambda}$  \\
\hline\hline
\end{longtable}

\begin{longtable}{llllll}
\caption{\label{p4pindeltab}Terms in the $\mathcal{O}(p^4)$ $\pi N\Delta$ chiral Lagrangian where $P^{(4)}_n$ is defined in Eq. \eqref{pindeln}.}\\
\hline\hline $n$ & $P^{(4)}_n$ & $n$ & $P^{(4)}_n$ & $n$ & $P^{(4)}_n$ \\
\hline\endfirsthead

\hline\hline $n$ & $P^{(4)}_n$ & $n$ & $P^{(4)}_n$ & $n$ & $P^{(4)}_n$ \\
\hline\endhead

\hline\hline 
\endfoot

\hline\endlastfoot

1 & $\psib u^{ i\mu}u_{ i\mu}u^{ j\nu}u^{ k\lambda}\tau_{ j}\gamf\gamma_{\nu}\psi_{A,n, k\lambda}$ & 74 & $\psib h^{ i\mu\nu}{{h^{ j}}_{\mu}}^{\lambda}\tau_{ i}\gamf\gamma_{\lambda}\psi_{A,n, j\nu}$ & 147 & $i\psib{f_{+}}^{ i\mu\nu}{{{f_{-}}^{ j}}_{\mu}}^{\lambda}\tau_{ j}D_{\lambda}\psi_{A,n, i\nu}$  \\
2 & $\psib u^{ i\mu}u_{ i\mu}u^{ j\nu}u^{ k\lambda}\tau_{ j}\gamf\gamma_{\lambda}\psi_{A,n, k\nu}$ & 75 & $\psib h^{ i\mu\nu}h^{ j\lambda\rho}\tau_{ i}\gamf\gamma_{\mu}D_{\nu\lambda}\psi_{A,n, j\rho}$ & 148 & $\psib{f_{+}}^{ i\mu\nu}{f_{-}}^{ j\lambda\rho}\tau_{ i}\sigma_{\mu\nu}D_{\lambda}\psi_{A,n, j\rho}$  \\
3 & $\psib u^{ i\mu}{u_{ i}}^{\nu}{u^{ j}}_{\mu}u^{ k\lambda}\tau_{ j}\gamf\gamma_{\nu}\psi_{A,n, k\lambda}$ & 76 & $\psib h^{ i\mu\nu}h^{ j\lambda\rho}\tau_{ i}\gamf\gamma_{\lambda}D_{\mu\nu}\psi_{A,n, j\rho}$ & 149 & $\psib{f_{+}}^{ i\mu\nu}{f_{-}}^{ j\lambda\rho}\tau_{ i}\sigma_{\mu\lambda}D_{\nu}\psi_{A,n, j\rho}$  \\
4 & $\psib u^{ i\mu}{u_{ i}}^{\nu}{u^{ j}}_{\mu}u^{ k\lambda}\tau_{ j}\gamf\gamma_{\lambda}\psi_{A,n, k\nu}$ & 77 & $\psib u^{ i\mu}\nabla_{\mu}{f_{-}}^{ j\nu\lambda}\tau_{ i}\gamf\gamma_{\nu}\psi_{A,n, j\lambda}$ & 150 & $\psib{f_{+}}^{ i\mu\nu}{f_{-}}^{ j\lambda\rho}\tau_{ j}\sigma_{\mu\nu}D_{\lambda}\psi_{A,n, i\rho}$  \\
5 & $\psib u^{ i\mu}{u_{ i}}^{\nu}{u^{ j}}_{\mu}u^{ k\lambda}\tau_{ k}\gamf\gamma_{\nu}\psi_{A,n, j\lambda}$ & 78 & $\psib u^{ i\mu}\nabla_{\mu}{f_{-}}^{ j\nu\lambda}\tau_{ j}\gamf\gamma_{\nu}\psi_{A,n, i\lambda}$ & 151 & $\psib{f_{+}}^{ i\mu\nu}{f_{-}}^{ j\lambda\rho}\tau_{ j}\sigma_{\mu\lambda}D_{\nu}\psi_{A,n, i\rho}$  \\
6 & $\psib u^{ i\mu}{u_{ i}}^{\nu}{u^{ j}}_{\mu}u^{ k\lambda}\tau_{ k}\gamf\gamma_{\lambda}\psi_{A,n, j\nu}$ & 79 & $\psib u^{ i\mu}\nabla^{\nu}{{{f_{-}}^{ j}}_{\mu}}^{\lambda}\tau_{ i}\gamf\gamma_{\nu}\psi_{A,n, j\lambda}$ & 152 & $i\psib{f_{+}}^{ i\mu\nu}{{h^{ j}}_{\mu}}^{\lambda}\tau_{ i}D_{\nu}\psi_{A,n, j\lambda}$  \\
\hline\hline
\end{longtable}

\bibliography{references}

\begin{thebibliography}{75}%
\makeatletter
\providecommand \@ifxundefined [1]{%
 \@ifx{#1\undefined}
}%
\providecommand \@ifnum [1]{%
 \ifnum #1\expandafter \@firstoftwo
 \else \expandafter \@secondoftwo
 \fi
}%
\providecommand \@ifx [1]{%
 \ifx #1\expandafter \@firstoftwo
 \else \expandafter \@secondoftwo
 \fi
}%
\providecommand \natexlab [1]{#1}%
\providecommand \enquote  [1]{``#1''}%
\providecommand \bibnamefont  [1]{#1}%
\providecommand \bibfnamefont [1]{#1}%
\providecommand \citenamefont [1]{#1}%
\providecommand \href@noop [0]{\@secondoftwo}%
\providecommand \href [0]{\begingroup \@sanitize@url \@href}%
\providecommand \@href[1]{\@@startlink{#1}\@@href}%
\providecommand \@@href[1]{\endgroup#1\@@endlink}%
\providecommand \@sanitize@url [0]{\catcode `\\12\catcode `\$12\catcode
  `\&12\catcode `\#12\catcode `\^12\catcode `\_12\catcode `\%12\relax}%
\providecommand \@@startlink[1]{}%
\providecommand \@@endlink[0]{}%
\providecommand \url  [0]{\begingroup\@sanitize@url \@url }%
\providecommand \@url [1]{\endgroup\@href {#1}{\urlprefix }}%
\providecommand \urlprefix  [0]{URL }%
\providecommand \Eprint [0]{\href }%
\providecommand \doibase [0]{http://dx.doi.org/}%
\providecommand \selectlanguage [0]{\@gobble}%
\providecommand \bibinfo  [0]{\@secondoftwo}%
\providecommand \bibfield  [0]{\@secondoftwo}%
\providecommand \translation [1]{[#1]}%
\providecommand \BibitemOpen [0]{}%
\providecommand \bibitemStop [0]{}%
\providecommand \bibitemNoStop [0]{.\EOS\space}%
\providecommand \EOS [0]{\spacefactor3000\relax}%
\providecommand \BibitemShut  [1]{\csname bibitem#1\endcsname}%
\let\auto@bib@innerbib\@empty
\bibitem [{\citenamefont {'t~Hooft}(1974)}]{tHooft:1973alw}%
  \BibitemOpen
  \bibfield  {author} {\bibinfo {author} {\bibfnamefont {Gerard}\ \bibnamefont
  {'t~Hooft}},\ }\bibfield  {title} {\enquote {\bibinfo {title} {{A Planar
  Diagram Theory for Strong Interactions}},}\ }\href {\doibase
  10.1016/0550-3213(74)90154-0} {\bibfield  {journal} {\bibinfo  {journal}
  {Nucl. Phys.}\ }\textbf {\bibinfo {volume} {B72}},\ \bibinfo {pages} {461}
  (\bibinfo {year} {1974})}\BibitemShut {NoStop}%
\bibitem [{\citenamefont {Witten}(1979)}]{Witten:1979kh}%
  \BibitemOpen
  \bibfield  {author} {\bibinfo {author} {\bibfnamefont {Edward}\ \bibnamefont
  {Witten}},\ }\bibfield  {title} {\enquote {\bibinfo {title} {{Baryons in the
  1/n Expansion}},}\ }\href {\doibase 10.1016/0550-3213(79)90232-3} {\bibfield
  {journal} {\bibinfo  {journal} {Nucl. Phys.}\ }\textbf {\bibinfo {volume}
  {B160}},\ \bibinfo {pages} {57--115} (\bibinfo {year} {1979})}\BibitemShut
  {NoStop}%
\bibitem [{\citenamefont {Weinberg}(1979)}]{weinberg}%
  \BibitemOpen
  \bibfield  {author} {\bibinfo {author} {\bibfnamefont {Steven}\ \bibnamefont
  {Weinberg}},\ }\bibfield  {title} {\enquote {\bibinfo {title}
  {{Phenomenological Lagrangians}},}\ }\href {\doibase
  10.1016/0378-4371(79)90223-1} {\bibfield  {journal} {\bibinfo  {journal}
  {Physica}\ }\textbf {\bibinfo {volume} {A 96}},\ \bibinfo {pages} {327--340}
  (\bibinfo {year} {1979})}\BibitemShut {NoStop}%
\bibitem [{\citenamefont {Gasser}\ and\ \citenamefont {Leutwyler}(1984)}]{GS1}%
  \BibitemOpen
  \bibfield  {author} {\bibinfo {author} {\bibfnamefont {J.}~\bibnamefont
  {Gasser}}\ and\ \bibinfo {author} {\bibfnamefont {H.}~\bibnamefont
  {Leutwyler}},\ }\bibfield  {title} {\enquote {\bibinfo {title} {{Chiral
  perturbation theory to one loop}},}\ }\href {\doibase
  10.1016/0003-4916(84)90242-2} {\bibfield  {journal} {\bibinfo  {journal}
  {Ann. Phys. (NY)}\ }\textbf {\bibinfo {volume} {158}},\ \bibinfo {pages}
  {142--210} (\bibinfo {year} {1984})}\BibitemShut {NoStop}%
\bibitem [{\citenamefont {Gasser}\ and\ \citenamefont {Leutwyler}(1985)}]{GS2}%
  \BibitemOpen
  \bibfield  {author} {\bibinfo {author} {\bibfnamefont {J.}~\bibnamefont
  {Gasser}}\ and\ \bibinfo {author} {\bibfnamefont {H.}~\bibnamefont
  {Leutwyler}},\ }\bibfield  {title} {\enquote {\bibinfo {title} {Chiral
  perturbation theory: Expansions in the mass of the strange quark},}\ }\href
  {\doibase 10.1016/0550-3213(85)90492-4} {\bibfield  {journal} {\bibinfo
  {journal} {Nucl. Phys. B}\ }\textbf {\bibinfo {volume} {250}},\ \bibinfo
  {pages} {465--516} (\bibinfo {year} {1985})}\BibitemShut {NoStop}%
\bibitem [{\citenamefont {Gasser}\ \emph {et~al.}(1988)\citenamefont {Gasser},
  \citenamefont {Sainio},\ and\ \citenamefont {\v{S}varc}}]{Gasser:1987rb}%
  \BibitemOpen
  \bibfield  {author} {\bibinfo {author} {\bibfnamefont {J.}~\bibnamefont
  {Gasser}}, \bibinfo {author} {\bibfnamefont {M.~E.}\ \bibnamefont {Sainio}},
  \ and\ \bibinfo {author} {\bibfnamefont {A.}~\bibnamefont {\v{S}varc}},\
  }\bibfield  {title} {\enquote {\bibinfo {title} {Nucleons with chiral
  loops},}\ }\href {\doibase 10.1016/0550-3213(88)90108-3} {\bibfield
  {journal} {\bibinfo  {journal} {Nucl. Phys. B}\ }\textbf {\bibinfo {volume}
  {307}},\ \bibinfo {pages} {779} (\bibinfo {year} {1988})}\BibitemShut
  {NoStop}%
\bibitem [{\citenamefont {Jenkins}\ and\ \citenamefont
  {Manohar}(1991{\natexlab{a}})}]{Jenkins:1991es}%
  \BibitemOpen
  \bibfield  {author} {\bibinfo {author} {\bibfnamefont {Elizabeth~Ellen}\
  \bibnamefont {Jenkins}}\ and\ \bibinfo {author} {\bibfnamefont {Aneesh~V.}\
  \bibnamefont {Manohar}},\ }\bibfield  {title} {\enquote {\bibinfo {title}
  {{Chiral corrections to the baryon axial currents}},}\ }\href {\doibase
  10.1016/0370-2693(91)90840-M} {\bibfield  {journal} {\bibinfo  {journal}
  {Phys. Lett.}\ }\textbf {\bibinfo {volume} {B259}},\ \bibinfo {pages}
  {353--358} (\bibinfo {year} {1991}{\natexlab{a}})}\BibitemShut {NoStop}%
\bibitem [{\citenamefont {Hemmert}\ \emph {et~al.}(1998)\citenamefont
  {Hemmert}, \citenamefont {Holstein},\ and\ \citenamefont
  {Kambor}}]{Hemmert:1997ye}%
  \BibitemOpen
  \bibfield  {author} {\bibinfo {author} {\bibfnamefont {Thomas~R.}\
  \bibnamefont {Hemmert}}, \bibinfo {author} {\bibfnamefont {Barry~R.}\
  \bibnamefont {Holstein}}, \ and\ \bibinfo {author} {\bibfnamefont {Joachim}\
  \bibnamefont {Kambor}},\ }\bibfield  {title} {\enquote {\bibinfo {title}
  {{Heavy baryon chiral perturbation theory with light deltas}},}\ }\href
  {\doibase 10.1088/0954-3899/24/10/003} {\bibfield  {journal} {\bibinfo
  {journal} {J. Phys. G}\ }\textbf {\bibinfo {volume} {24}},\ \bibinfo {pages}
  {1831--1859} (\bibinfo {year} {1998})},\ \Eprint
  {http://arxiv.org/abs/hep-ph/9712496} {arXiv:hep-ph/9712496 [hep-ph]}
  \BibitemShut {NoStop}%
\bibitem [{\citenamefont {Jenkins}\ and\ \citenamefont
  {Manohar}(1991{\natexlab{b}})}]{Jenkins:1990jv}%
  \BibitemOpen
  \bibfield  {author} {\bibinfo {author} {\bibfnamefont {Elizabeth~Ellen}\
  \bibnamefont {Jenkins}}\ and\ \bibinfo {author} {\bibfnamefont {Aneesh~V.}\
  \bibnamefont {Manohar}},\ }\bibfield  {title} {\enquote {\bibinfo {title}
  {{Baryon chiral perturbation theory using a heavy fermion Lagrangian}},}\
  }\href {\doibase 10.1016/0370-2693(91)90266-S} {\bibfield  {journal}
  {\bibinfo  {journal} {Phys. Lett.}\ }\textbf {\bibinfo {volume} {B255}},\
  \bibinfo {pages} {558--562} (\bibinfo {year}
  {1991}{\natexlab{b}})}\BibitemShut {NoStop}%
\bibitem [{\citenamefont {Becher}\ and\ \citenamefont
  {Leutwyler}(1999)}]{Becher:1999he}%
  \BibitemOpen
  \bibfield  {author} {\bibinfo {author} {\bibfnamefont {Thomas}\ \bibnamefont
  {Becher}}\ and\ \bibinfo {author} {\bibfnamefont {H.}~\bibnamefont
  {Leutwyler}},\ }\bibfield  {title} {\enquote {\bibinfo {title} {{Baryon
  chiral perturbation theory in manifestly Lorentz invariant form}},}\ }\href
  {\doibase 10.1007/PL00021673} {\bibfield  {journal} {\bibinfo  {journal}
  {Eur. Phys. J.}\ }\textbf {\bibinfo {volume} {C9}},\ \bibinfo {pages}
  {643--671} (\bibinfo {year} {1999})},\ \Eprint
  {http://arxiv.org/abs/hep-ph/9901384} {arXiv:hep-ph/9901384 [hep-ph]}
  \BibitemShut {NoStop}%
\bibitem [{\citenamefont {Fuchs}\ \emph {et~al.}(2003)\citenamefont {Fuchs},
  \citenamefont {Gegelia}, \citenamefont {Japaridze},\ and\ \citenamefont
  {Scherer}}]{Fuchs:2003qc}%
  \BibitemOpen
  \bibfield  {author} {\bibinfo {author} {\bibfnamefont {T.}~\bibnamefont
  {Fuchs}}, \bibinfo {author} {\bibfnamefont {J.}~\bibnamefont {Gegelia}},
  \bibinfo {author} {\bibfnamefont {G.}~\bibnamefont {Japaridze}}, \ and\
  \bibinfo {author} {\bibfnamefont {S.}~\bibnamefont {Scherer}},\ }\bibfield
  {title} {\enquote {\bibinfo {title} {{Renormalization of relativistic baryon
  chiral perturbation theory and power counting}},}\ }\href {\doibase
  10.1103/PhysRevD.68.056005} {\bibfield  {journal} {\bibinfo  {journal} {Phys.
  Rev.}\ }\textbf {\bibinfo {volume} {D68}},\ \bibinfo {pages} {056005}
  (\bibinfo {year} {2003})},\ \Eprint {http://arxiv.org/abs/hep-ph/0302117}
  {arXiv:hep-ph/0302117 [hep-ph]} \BibitemShut {NoStop}%
\bibitem [{\citenamefont {Pascalutsa}\ and\ \citenamefont
  {Phillips}(2003)}]{Pascalutsa:2002pi}%
  \BibitemOpen
  \bibfield  {author} {\bibinfo {author} {\bibfnamefont {Vladimir}\
  \bibnamefont {Pascalutsa}}\ and\ \bibinfo {author} {\bibfnamefont
  {Daniel~R.}\ \bibnamefont {Phillips}},\ }\bibfield  {title} {\enquote
  {\bibinfo {title} {{Effective theory of the delta(1232) in Compton scattering
  off the nucleon}},}\ }\href {\doibase 10.1103/PhysRevC.67.055202} {\bibfield
  {journal} {\bibinfo  {journal} {Phys. Rev.}\ }\textbf {\bibinfo {volume}
  {C67}},\ \bibinfo {pages} {055202} (\bibinfo {year} {2003})},\ \Eprint
  {http://arxiv.org/abs/nucl-th/0212024} {arXiv:nucl-th/0212024 [nucl-th]}
  \BibitemShut {NoStop}%
\bibitem [{\citenamefont {Fearing}\ and\ \citenamefont {Scherer}(1996)}]{p61}%
  \BibitemOpen
  \bibfield  {author} {\bibinfo {author} {\bibfnamefont {H.~W.}\ \bibnamefont
  {Fearing}}\ and\ \bibinfo {author} {\bibfnamefont {S.}~\bibnamefont
  {Scherer}},\ }\bibfield  {title} {\enquote {\bibinfo {title} {Extension of
  the chiral perturbation theory meson {Lagrangian} to order $p^6$},}\ }\href
  {\doibase 10.1103/PhysRevD.53.315} {\bibfield  {journal} {\bibinfo  {journal}
  {Phys. Rev. D}\ }\textbf {\bibinfo {volume} {53}},\ \bibinfo {pages}
  {315--348} (\bibinfo {year} {1996})},\ \Eprint
  {http://arxiv.org/abs/hep-ph/9408346} {arXiv:hep-ph/9408346 [hep-ph]}
  \BibitemShut {NoStop}%
\bibitem [{\citenamefont {Bijnens}\ \emph {et~al.}(1999)\citenamefont
  {Bijnens}, \citenamefont {Colangelo},\ and\ \citenamefont {Ecker}}]{p62}%
  \BibitemOpen
  \bibfield  {author} {\bibinfo {author} {\bibfnamefont {Johan}\ \bibnamefont
  {Bijnens}}, \bibinfo {author} {\bibfnamefont {Gilberto}\ \bibnamefont
  {Colangelo}}, \ and\ \bibinfo {author} {\bibfnamefont {Gerhard}\ \bibnamefont
  {Ecker}},\ }\bibfield  {title} {\enquote {\bibinfo {title} {{The mesonic
  chiral Lagrangian of order $p^6$}},}\ }\href {\doibase
  10.1088/1126-6708/1999/02/020} {\bibfield  {journal} {\bibinfo  {journal}
  {Journal of High Energy Physics}\ }\textbf {\bibinfo {volume} {9902}},\
  \bibinfo {pages} {020} (\bibinfo {year} {1999})},\ \Eprint
  {http://arxiv.org/abs/hep-ph/9902437} {arXiv:hep-ph/9902437 [hep-ph]}
  \BibitemShut {NoStop}%
\bibitem [{\citenamefont {Haefeli}\ \emph {et~al.}(2007)\citenamefont
  {Haefeli}, \citenamefont {Ivanov}, \citenamefont {Schmid},\ and\
  \citenamefont {Ecker}}]{p6p}%
  \BibitemOpen
  \bibfield  {author} {\bibinfo {author} {\bibfnamefont {Christoph}\
  \bibnamefont {Haefeli}}, \bibinfo {author} {\bibfnamefont {Mikhail~A.}\
  \bibnamefont {Ivanov}}, \bibinfo {author} {\bibfnamefont {Martin}\
  \bibnamefont {Schmid}}, \ and\ \bibinfo {author} {\bibfnamefont {Gerhard}\
  \bibnamefont {Ecker}},\ }\bibfield  {title} {\enquote {\bibinfo {title} {On
  the mesonic {Lagrangian} of order $p^6$ in chiral {SU(2)}},}\ }\href@noop {}
  {\  (\bibinfo {year} {2007})},\ \Eprint {http://arxiv.org/abs/0705.0576}
  {arXiv:0705.0576 [hep-ph]} \BibitemShut {NoStop}%
\bibitem [{\citenamefont {Ebertsh{\"{a}}user}\ \emph
  {et~al.}(2002)\citenamefont {Ebertsh{\"{a}}user}, \citenamefont {Fearing},\
  and\ \citenamefont {Scherer}}]{p6a1}%
  \BibitemOpen
  \bibfield  {author} {\bibinfo {author} {\bibfnamefont {T.}~\bibnamefont
  {Ebertsh{\"{a}}user}}, \bibinfo {author} {\bibfnamefont {H.~W.}\ \bibnamefont
  {Fearing}}, \ and\ \bibinfo {author} {\bibfnamefont {S.}~\bibnamefont
  {Scherer}},\ }\bibfield  {title} {\enquote {\bibinfo {title} {The anomalous
  chiral perturbation theory meson {Lagrangian} to order $p^6$ reexamined},}\
  }\href {\doibase 10.1103/PhysRevD.65.054033} {\bibfield  {journal} {\bibinfo
  {journal} {Phys. Rev. D}\ }\textbf {\bibinfo {volume} {65}},\ \bibinfo
  {pages} {054033} (\bibinfo {year} {2002})},\ \Eprint
  {http://arxiv.org/abs/hep-ph/0110261} {arXiv:hep-ph/0110261 [hep-ph]}
  \BibitemShut {NoStop}%
\bibitem [{\citenamefont {Bijnens}\ \emph {et~al.}(2002)\citenamefont
  {Bijnens}, \citenamefont {Girlanda},\ and\ \citenamefont {Talavera}}]{p6a2}%
  \BibitemOpen
  \bibfield  {author} {\bibinfo {author} {\bibfnamefont {J.}~\bibnamefont
  {Bijnens}}, \bibinfo {author} {\bibfnamefont {L.}~\bibnamefont {Girlanda}}, \
  and\ \bibinfo {author} {\bibfnamefont {P.}~\bibnamefont {Talavera}},\
  }\bibfield  {title} {\enquote {\bibinfo {title} {{The anomalous chiral
  Lagrangian of order $p^6$}},}\ }\href {\doibase 10.1007/s100520100887}
  {\bibfield  {journal} {\bibinfo  {journal} {Eur. Phys. J. C}\ }\textbf
  {\bibinfo {volume} {23}},\ \bibinfo {pages} {539--544} (\bibinfo {year}
  {2002})},\ \Eprint {http://arxiv.org/abs/hep-ph/0110400}
  {arXiv:hep-ph/0110400 [hep-ph]} \BibitemShut {NoStop}%
\bibitem [{\citenamefont {Cat\`a}\ and\ \citenamefont {Mateu}(2007)}]{tensor1}%
  \BibitemOpen
  \bibfield  {author} {\bibinfo {author} {\bibfnamefont {Oscar}\ \bibnamefont
  {Cat\`a}}\ and\ \bibinfo {author} {\bibfnamefont {Vicent}\ \bibnamefont
  {Mateu}},\ }\bibfield  {title} {\enquote {\bibinfo {title} {Chiral
  perturbation theory with tensor sources},}\ }\href {\doibase
  10.1088/1126-6708/2007/09/078} {\bibfield  {journal} {\bibinfo  {journal}
  {Journal of High Energy Physics}\ }\textbf {\bibinfo {volume} {0709}},\
  \bibinfo {pages} {078} (\bibinfo {year} {2007})},\ \Eprint
  {http://arxiv.org/abs/0705.2948} {arXiv:0705.2948 [hep-ph]} \BibitemShut
  {NoStop}%
\bibitem [{\citenamefont {Herrera-Sikl\'ody}\ \emph {et~al.}(1997)\citenamefont
  {Herrera-Sikl\'ody}, \citenamefont {Latorre}, \citenamefont {Pascual},\ and\
  \citenamefont {Taron}}]{U3}%
  \BibitemOpen
  \bibfield  {author} {\bibinfo {author} {\bibfnamefont {P.}~\bibnamefont
  {Herrera-Sikl\'ody}}, \bibinfo {author} {\bibfnamefont {J.I.}\ \bibnamefont
  {Latorre}}, \bibinfo {author} {\bibfnamefont {P.}~\bibnamefont {Pascual}}, \
  and\ \bibinfo {author} {\bibfnamefont {J.}~\bibnamefont {Taron}},\ }\bibfield
   {title} {\enquote {\bibinfo {title} {Chiral effective lagrangian in the
  large $n_c$ limit: the nonet case},}\ }\href {\doibase
  10.1016/S0550-3213(97)00260-5} {\bibfield  {journal} {\bibinfo  {journal}
  {Nucl. Phys. B}\ }\textbf {\bibinfo {volume} {497}},\ \bibinfo {pages}
  {345--386} (\bibinfo {year} {1997})},\ \Eprint
  {http://arxiv.org/abs/hep-ph/9610549} {arXiv:hep-ph/9610549 [hep-ph]}
  \BibitemShut {NoStop}%
\bibitem [{\citenamefont {Jiang}\ \emph {et~al.}(2014)\citenamefont {Jiang},
  \citenamefont {Ge},\ and\ \citenamefont {Wang}}]{ourf}%
  \BibitemOpen
  \bibfield  {author} {\bibinfo {author} {\bibfnamefont {Shao-Zhou}\
  \bibnamefont {Jiang}}, \bibinfo {author} {\bibfnamefont {Feng-Jun}\
  \bibnamefont {Ge}}, \ and\ \bibinfo {author} {\bibfnamefont {Qing}\
  \bibnamefont {Wang}},\ }\bibfield  {title} {\enquote {\bibinfo {title} {Full
  pseudoscalar mesonic chiral lagrangian at $p^6$ order under the unitary
  group},}\ }\href {\doibase 10.1103/PhysRevD.89.074048} {\bibfield  {journal}
  {\bibinfo  {journal} {Phys. Rev. D}\ }\textbf {\bibinfo {volume} {89}},\
  \bibinfo {pages} {074048} (\bibinfo {year} {2014})},\ \Eprint
  {http://arxiv.org/abs/1401.0317} {arXiv:1401.0317 [hep-ph]} \BibitemShut
  {NoStop}%
\bibitem [{\citenamefont {Krause}(1990)}]{Krause:1990xc}%
  \BibitemOpen
  \bibfield  {author} {\bibinfo {author} {\bibfnamefont {Andreas}\ \bibnamefont
  {Krause}},\ }\bibfield  {title} {\enquote {\bibinfo {title} {Baryon matrix
  elements of the vector current in chiral perturbation theory},}\ }\href
  {\doibase 10.5169/seals-116214} {\bibfield  {journal} {\bibinfo  {journal}
  {Helv. Phys. Acta}\ }\textbf {\bibinfo {volume} {63}},\ \bibinfo {pages}
  {3--70} (\bibinfo {year} {1990})}\BibitemShut {NoStop}%
\bibitem [{\citenamefont {Ecker}(1994)}]{Ecker:1994pi}%
  \BibitemOpen
  \bibfield  {author} {\bibinfo {author} {\bibfnamefont {G.}~\bibnamefont
  {Ecker}},\ }\bibfield  {title} {\enquote {\bibinfo {title} {Chiral invariant
  renormalization of the pion-nucleon interaction},}\ }\href {\doibase
  10.1016/0370-2693(94)90565-7} {\bibfield  {journal} {\bibinfo  {journal}
  {Phys. Lett. B}\ }\textbf {\bibinfo {volume} {336}},\ \bibinfo {pages}
  {508--517} (\bibinfo {year} {1994})},\ \Eprint
  {http://arxiv.org/abs/hep-ph/9402337} {arXiv:hep-ph/9402337 [hep-ph]}
  \BibitemShut {NoStop}%
\bibitem [{\citenamefont {{Fettes}}\ \emph {et~al.}(1998)\citenamefont
  {{Fettes}}, \citenamefont {{Mei{\ss}ner}},\ and\ \citenamefont
  {{Steininger}}}]{1998NuPhA.640..199F}%
  \BibitemOpen
  \bibfield  {author} {\bibinfo {author} {\bibfnamefont {Nadia}\ \bibnamefont
  {{Fettes}}}, \bibinfo {author} {\bibfnamefont {Ulf-G.}\ \bibnamefont
  {{Mei{\ss}ner}}}, \ and\ \bibinfo {author} {\bibfnamefont {Sven}\
  \bibnamefont {{Steininger}}},\ }\bibfield  {title} {\enquote {\bibinfo
  {title} {Pion-nucleon scattering in chiral perturbation theory (i):
  Isospin-symmetric case},}\ }\href {\doibase 10.1016/S0375-9474(98)00452-7}
  {\bibfield  {journal} {\bibinfo  {journal} {Nucl. Phys. A}\ }\textbf
  {\bibinfo {volume} {640}},\ \bibinfo {pages} {199--234} (\bibinfo {year}
  {1998})},\ \Eprint {http://arxiv.org/abs/hep-ph/9803266} {hep-ph/9803266}
  \BibitemShut {NoStop}%
\bibitem [{\citenamefont {Mei{\ss}ner}\ \emph {et~al.}(2000)\citenamefont
  {Mei{\ss}ner}, \citenamefont {Muller},\ and\ \citenamefont
  {Steininger}}]{Meissner:1998rw}%
  \BibitemOpen
  \bibfield  {author} {\bibinfo {author} {\bibfnamefont {Ulf-G.}\ \bibnamefont
  {Mei{\ss}ner}}, \bibinfo {author} {\bibfnamefont {Guido}\ \bibnamefont
  {Muller}}, \ and\ \bibinfo {author} {\bibfnamefont {Sven}\ \bibnamefont
  {Steininger}},\ }\bibfield  {title} {\enquote {\bibinfo {title}
  {Renormalization of the chiral pion - nucleon lagrangian beyond
  next-to-leading order},}\ }\href {\doibase 10.1006/aphy.1999.5919} {\bibfield
   {journal} {\bibinfo  {journal} {Ann. Phys. (NY)}\ }\textbf {\bibinfo
  {volume} {279}},\ \bibinfo {pages} {1--64} (\bibinfo {year} {2000})},\
  \Eprint {http://arxiv.org/abs/hep-ph/9809446} {arXiv:hep-ph/9809446 [hep-ph]}
  \BibitemShut {NoStop}%
\bibitem [{\citenamefont {{Fettes}}\ \emph {et~al.}(2000)\citenamefont
  {{Fettes}}, \citenamefont {{Mei{\ss}ner}}, \citenamefont {{Moj{\v z}i{\v
  s}}},\ and\ \citenamefont {{Steininger}}}]{pin4}%
  \BibitemOpen
  \bibfield  {author} {\bibinfo {author} {\bibfnamefont {Nadia}\ \bibnamefont
  {{Fettes}}}, \bibinfo {author} {\bibfnamefont {Ulf-G.}\ \bibnamefont
  {{Mei{\ss}ner}}}, \bibinfo {author} {\bibfnamefont {Martin}\ \bibnamefont
  {{Moj{\v z}i{\v s}}}}, \ and\ \bibinfo {author} {\bibfnamefont {Sven}\
  \bibnamefont {{Steininger}}},\ }\bibfield  {title} {\enquote {\bibinfo
  {title} {The chiral effective pion nucleon lagrangian of order $p^4$},}\
  }\href {\doibase 10.1006/aphy.2000.6059} {\bibfield  {journal} {\bibinfo
  {journal} {Ann. Phys. (NY)}\ }\textbf {\bibinfo {volume} {283}},\ \bibinfo
  {pages} {273--302} (\bibinfo {year} {2000})},\ \Eprint
  {http://arxiv.org/abs/hep-ph/0001308} {arXiv:hep-ph/0001308 [hep-ph]}
  \BibitemShut {NoStop}%
\bibitem [{\citenamefont {Oller}\ \emph {et~al.}(2006)\citenamefont {Oller},
  \citenamefont {Verbeni},\ and\ \citenamefont {Prades}}]{pib31}%
  \BibitemOpen
  \bibfield  {author} {\bibinfo {author} {\bibfnamefont {Jos\'{e}~Antonio}\
  \bibnamefont {Oller}}, \bibinfo {author} {\bibfnamefont {Michela}\
  \bibnamefont {Verbeni}}, \ and\ \bibinfo {author} {\bibfnamefont {Joaquim}\
  \bibnamefont {Prades}},\ }\bibfield  {title} {\enquote {\bibinfo {title}
  {Meson-baryon effective chiral lagrangians to $\mathcal{O}(q^3)$},}\ }\href
  {\doibase 10.1088/1126-6708/2006/09/079} {\bibfield  {journal} {\bibinfo
  {journal} {Journal of High Energy Physics}\ }\textbf {\bibinfo {volume}
  {0609}},\ \bibinfo {pages} {079} (\bibinfo {year} {2006})},\ \Eprint
  {http://arxiv.org/abs/hep-ph/0608204} {arXiv:hep-ph/0608204 [hep-ph]}
  \BibitemShut {NoStop}%
\bibitem [{\citenamefont {Frink}\ and\ \citenamefont
  {Mei{\ss}ner}(2006)}]{pib32}%
  \BibitemOpen
  \bibfield  {author} {\bibinfo {author} {\bibfnamefont {M.}~\bibnamefont
  {Frink}}\ and\ \bibinfo {author} {\bibfnamefont {U.-G.}\ \bibnamefont
  {Mei{\ss}ner}},\ }\bibfield  {title} {\enquote {\bibinfo {title} {On the
  chiral effective meson-baryon lagrangian at third order},}\ }\href {\doibase
  10.1140/epja/i2006-10105-x} {\bibfield  {journal} {\bibinfo  {journal} {Eur.
  Phys. J. A}\ }\textbf {\bibinfo {volume} {29}},\ \bibinfo {pages} {255--260}
  (\bibinfo {year} {2006})},\ \Eprint {http://arxiv.org/abs/hep-ph/0609256}
  {arXiv:hep-ph/0609256 [hep-ph]} \BibitemShut {NoStop}%
\bibitem [{\citenamefont {Jiang}\ \emph {et~al.}(2017)\citenamefont {Jiang},
  \citenamefont {Chen},\ and\ \citenamefont {Liu}}]{Jiang:2016vax}%
  \BibitemOpen
  \bibfield  {author} {\bibinfo {author} {\bibfnamefont {Shao-Zhou}\
  \bibnamefont {Jiang}}, \bibinfo {author} {\bibfnamefont {Qing-Sen}\
  \bibnamefont {Chen}}, \ and\ \bibinfo {author} {\bibfnamefont {Yan-Rui}\
  \bibnamefont {Liu}},\ }\bibfield  {title} {\enquote {\bibinfo {title}
  {{Meson-baryon effective chiral Lagrangians at order $p^4$}},}\ }\href
  {\doibase 10.1103/PhysRevD.95.014012} {\bibfield  {journal} {\bibinfo
  {journal} {Phys. Rev. D}\ }\textbf {\bibinfo {volume} {95}},\ \bibinfo
  {pages} {014012} (\bibinfo {year} {2017})},\ \Eprint
  {http://arxiv.org/abs/1608.06104} {arXiv:1608.06104 [hep-ph]} \BibitemShut
  {NoStop}%
\bibitem [{\citenamefont {Pascalutsa}\ and\ \citenamefont
  {Timmermans}(1999)}]{Pascalutsa:1999zz}%
  \BibitemOpen
  \bibfield  {author} {\bibinfo {author} {\bibfnamefont {Vladimir}\
  \bibnamefont {Pascalutsa}}\ and\ \bibinfo {author} {\bibfnamefont {Rob}\
  \bibnamefont {Timmermans}},\ }\bibfield  {title} {\enquote {\bibinfo {title}
  {{Field theory of nucleon to higher spin baryon transitions}},}\ }\href
  {\doibase 10.1103/PhysRevC.60.042201} {\bibfield  {journal} {\bibinfo
  {journal} {Phys. Rev. C}\ }\textbf {\bibinfo {volume} {60}},\ \bibinfo
  {pages} {042201} (\bibinfo {year} {1999})},\ \Eprint
  {http://arxiv.org/abs/nucl-th/9905065} {arXiv:nucl-th/9905065 [nucl-th]}
  \BibitemShut {NoStop}%
\bibitem [{\citenamefont {Pascalutsa}\ and\ \citenamefont
  {Vanderhaeghen}(2005{\natexlab{a}})}]{Pascalutsa:2005ts}%
  \BibitemOpen
  \bibfield  {author} {\bibinfo {author} {\bibfnamefont {Vladimir}\
  \bibnamefont {Pascalutsa}}\ and\ \bibinfo {author} {\bibfnamefont {Marc}\
  \bibnamefont {Vanderhaeghen}},\ }\bibfield  {title} {\enquote {\bibinfo
  {title} {{Electromagnetic nucleon-to-Delta transition in chiral
  effective-field theory}},}\ }\href {\doibase 10.1103/PhysRevLett.95.232001}
  {\bibfield  {journal} {\bibinfo  {journal} {Phys. Rev. Lett.}\ }\textbf
  {\bibinfo {volume} {95}},\ \bibinfo {pages} {232001} (\bibinfo {year}
  {2005}{\natexlab{a}})},\ \Eprint {http://arxiv.org/abs/hep-ph/0508060}
  {arXiv:hep-ph/0508060 [hep-ph]} \BibitemShut {NoStop}%
\bibitem [{\citenamefont {Procura}(2008)}]{Procura:2008ze}%
  \BibitemOpen
  \bibfield  {author} {\bibinfo {author} {\bibfnamefont {Massimiliano}\
  \bibnamefont {Procura}},\ }\bibfield  {title} {\enquote {\bibinfo {title}
  {{Chiral symmetry and the axial nucleon to $\Delta$(1232) transition form
  factors}},}\ }\href {\doibase 10.1103/PhysRevD.78.094021} {\bibfield
  {journal} {\bibinfo  {journal} {Phys. Rev. D}\ }\textbf {\bibinfo {volume}
  {78}},\ \bibinfo {pages} {094021} (\bibinfo {year} {2008})},\ \Eprint
  {http://arxiv.org/abs/0803.4291} {arXiv:0803.4291 [hep-ph]} \BibitemShut
  {NoStop}%
\bibitem [{\citenamefont {Li}\ \emph {et~al.}(2017)\citenamefont {Li},
  \citenamefont {Liu}, \citenamefont {Chen}, \citenamefont {Deng},\ and\
  \citenamefont {Zhu}}]{Li:2017vmq}%
  \BibitemOpen
  \bibfield  {author} {\bibinfo {author} {\bibfnamefont {Hao-Song}\
  \bibnamefont {Li}}, \bibinfo {author} {\bibfnamefont {Zhan-Wei}\ \bibnamefont
  {Liu}}, \bibinfo {author} {\bibfnamefont {Xiao-Lin}\ \bibnamefont {Chen}},
  \bibinfo {author} {\bibfnamefont {Wei-Zhen}\ \bibnamefont {Deng}}, \ and\
  \bibinfo {author} {\bibfnamefont {Shi-Lin}\ \bibnamefont {Zhu}},\ }\bibfield
  {title} {\enquote {\bibinfo {title} {{Decuplet to octet baryon transitions in
  chiral perturbation theory}},}\ }\href@noop {} {\  (\bibinfo {year}
  {2017})},\ \Eprint {http://arxiv.org/abs/1706.06458} {arXiv:1706.06458
  [hep-ph]} \BibitemShut {NoStop}%
\bibitem [{\citenamefont {Gellas}\ \emph {et~al.}(1999)\citenamefont {Gellas},
  \citenamefont {Hemmert}, \citenamefont {Ktorides},\ and\ \citenamefont
  {Poulis}}]{Gellas:1998wx}%
  \BibitemOpen
  \bibfield  {author} {\bibinfo {author} {\bibfnamefont {G.~C.}\ \bibnamefont
  {Gellas}}, \bibinfo {author} {\bibfnamefont {T.~R.}\ \bibnamefont {Hemmert}},
  \bibinfo {author} {\bibfnamefont {C.~N.}\ \bibnamefont {Ktorides}}, \ and\
  \bibinfo {author} {\bibfnamefont {G.~I.}\ \bibnamefont {Poulis}},\ }\bibfield
   {title} {\enquote {\bibinfo {title} {{Chiral symmetry and the delta-nucleon
  transition form factors}},}\ }\href {\doibase 10.1103/PhysRevD.60.054022}
  {\bibfield  {journal} {\bibinfo  {journal} {Phys. Rev. D}\ }\textbf {\bibinfo
  {volume} {60}},\ \bibinfo {pages} {054022} (\bibinfo {year} {1999})},\
  \Eprint {http://arxiv.org/abs/hep-ph/9810426} {arXiv:hep-ph/9810426 [hep-ph]}
  \BibitemShut {NoStop}%
\bibitem [{\citenamefont {Pascalutsa}\ and\ \citenamefont
  {Vanderhaeghen}(2005{\natexlab{b}})}]{Pascalutsa:2004je}%
  \BibitemOpen
  \bibfield  {author} {\bibinfo {author} {\bibfnamefont {Vladimir}\
  \bibnamefont {Pascalutsa}}\ and\ \bibinfo {author} {\bibfnamefont {Marc}\
  \bibnamefont {Vanderhaeghen}},\ }\bibfield  {title} {\enquote {\bibinfo
  {title} {{Magnetic moment of the $\Delta(1232)$ resonance in chiral effective
  field theory}},}\ }\href {\doibase 10.1103/PhysRevLett.94.102003} {\bibfield
  {journal} {\bibinfo  {journal} {Phys. Rev. Lett.}\ }\textbf {\bibinfo
  {volume} {94}},\ \bibinfo {pages} {102003} (\bibinfo {year}
  {2005}{\natexlab{b}})},\ \Eprint {http://arxiv.org/abs/nucl-th/0412113}
  {arXiv:nucl-th/0412113 [nucl-th]} \BibitemShut {NoStop}%
\bibitem [{\citenamefont {Pascalutsa}\ \emph {et~al.}(2007)\citenamefont
  {Pascalutsa}, \citenamefont {Vanderhaeghen},\ and\ \citenamefont
  {Yang}}]{Pascalutsa:2006up}%
  \BibitemOpen
  \bibfield  {author} {\bibinfo {author} {\bibfnamefont {Vladimir}\
  \bibnamefont {Pascalutsa}}, \bibinfo {author} {\bibfnamefont {Marc}\
  \bibnamefont {Vanderhaeghen}}, \ and\ \bibinfo {author} {\bibfnamefont
  {Shin~Nan}\ \bibnamefont {Yang}},\ }\bibfield  {title} {\enquote {\bibinfo
  {title} {{Electromagnetic excitation of the $\Delta$(1232)-resonance}},}\
  }\href {\doibase 10.1016/j.physrep.2006.09.006} {\bibfield  {journal}
  {\bibinfo  {journal} {Phys. Rept.}\ }\textbf {\bibinfo {volume} {437}},\
  \bibinfo {pages} {125--232} (\bibinfo {year} {2007})},\ \Eprint
  {http://arxiv.org/abs/hep-ph/0609004} {arXiv:hep-ph/0609004 [hep-ph]}
  \BibitemShut {NoStop}%
\bibitem [{\citenamefont {Pascalutsa}(2008)}]{Pascalutsa:2007yg}%
  \BibitemOpen
  \bibfield  {author} {\bibinfo {author} {\bibfnamefont {Vladimir}\
  \bibnamefont {Pascalutsa}},\ }\bibfield  {title} {\enquote {\bibinfo {title}
  {{The $\Delta$(1232)-resonance in chiral effective field theory}},}\ }\href
  {\doibase 10.1016/j.ppnp.2007.12.023} {\bibfield  {journal} {\bibinfo
  {journal} {Prog. Part. Nucl. Phys.}\ }\textbf {\bibinfo {volume} {61}},\
  \bibinfo {pages} {27--33} (\bibinfo {year} {2008})},\ \Eprint
  {http://arxiv.org/abs/0712.3919} {arXiv:0712.3919 [nucl-th]} \BibitemShut
  {NoStop}%
\bibitem [{\citenamefont {Pascalutsa}\ and\ \citenamefont
  {Vanderhaeghen}(2008)}]{Pascalutsa:2007wb}%
  \BibitemOpen
  \bibfield  {author} {\bibinfo {author} {\bibfnamefont {Vladimir}\
  \bibnamefont {Pascalutsa}}\ and\ \bibinfo {author} {\bibfnamefont {Marc}\
  \bibnamefont {Vanderhaeghen}},\ }\bibfield  {title} {\enquote {\bibinfo
  {title} {{Chiral effective-field theory in the $\Delta$(1232) region. II.
  Radiative pion photoproduction}},}\ }\href {\doibase
  10.1103/PhysRevD.77.014027} {\bibfield  {journal} {\bibinfo  {journal} {Phys.
  Rev. D}\ }\textbf {\bibinfo {volume} {77}},\ \bibinfo {pages} {014027}
  (\bibinfo {year} {2008})},\ \Eprint {http://arxiv.org/abs/0709.4583}
  {arXiv:0709.4583 [hep-ph]} \BibitemShut {NoStop}%
\bibitem [{\citenamefont {Bernard}\ \emph {et~al.}(2008)\citenamefont
  {Bernard}, \citenamefont {Mei\ss{}ner},\ and\ \citenamefont
  {Rusetsky}}]{Bernard:2007cm}%
  \BibitemOpen
  \bibfield  {author} {\bibinfo {author} {\bibfnamefont {V\'{e}ronique}\
  \bibnamefont {Bernard}}, \bibinfo {author} {\bibfnamefont {Ulf-G.}\
  \bibnamefont {Mei\ss{}ner}}, \ and\ \bibinfo {author} {\bibfnamefont {Akaki}\
  \bibnamefont {Rusetsky}},\ }\bibfield  {title} {\enquote {\bibinfo {title}
  {{The $\Delta$-resonance in a finite volume}},}\ }\href {\doibase
  10.1016/j.nuclphysb.2007.07.030} {\bibfield  {journal} {\bibinfo  {journal}
  {Nucl. Phys. B}\ }\textbf {\bibinfo {volume} {788}},\ \bibinfo {pages}
  {1--20} (\bibinfo {year} {2008})},\ \Eprint
  {http://arxiv.org/abs/hep-lat/0702012} {arXiv:hep-lat/0702012 [HEP-LAT]}
  \BibitemShut {NoStop}%
\bibitem [{\citenamefont {Bernard}\ \emph {et~al.}(2009)\citenamefont
  {Bernard}, \citenamefont {Hoja}, \citenamefont {Meissner},\ and\
  \citenamefont {Rusetsky}}]{Bernard:2009mw}%
  \BibitemOpen
  \bibfield  {author} {\bibinfo {author} {\bibfnamefont {V.}~\bibnamefont
  {Bernard}}, \bibinfo {author} {\bibfnamefont {D.}~\bibnamefont {Hoja}},
  \bibinfo {author} {\bibfnamefont {U.~G.}\ \bibnamefont {Meissner}}, \ and\
  \bibinfo {author} {\bibfnamefont {A.}~\bibnamefont {Rusetsky}},\ }\bibfield
  {title} {\enquote {\bibinfo {title} {{The Mass of the Delta resonance in a
  finite volume: fourth-order calculation}},}\ }\href {\doibase
  10.1088/1126-6708/2009/06/061} {\bibfield  {journal} {\bibinfo  {journal}
  {Journal of High Energy Physics}\ }\textbf {\bibinfo {volume} {06}},\
  \bibinfo {pages} {061} (\bibinfo {year} {2009})},\ \Eprint
  {http://arxiv.org/abs/0902.2346} {arXiv:0902.2346 [hep-lat]} \BibitemShut
  {NoStop}%
\bibitem [{\citenamefont {Ren}\ \emph {et~al.}(2014)\citenamefont {Ren},
  \citenamefont {Geng},\ and\ \citenamefont {Meng}}]{Ren:2013oaa}%
  \BibitemOpen
  \bibfield  {author} {\bibinfo {author} {\bibfnamefont {Xiu-Lei}\ \bibnamefont
  {Ren}}, \bibinfo {author} {\bibfnamefont {Li-Sheng}\ \bibnamefont {Geng}}, \
  and\ \bibinfo {author} {\bibfnamefont {Jie}\ \bibnamefont {Meng}},\
  }\bibfield  {title} {\enquote {\bibinfo {title} {{Decuplet baryon masses in
  covariant baryon chiral perturbation theory}},}\ }\href {\doibase
  10.1103/PhysRevD.89.054034} {\bibfield  {journal} {\bibinfo  {journal} {Phys.
  Rev.}\ }\textbf {\bibinfo {volume} {D89}},\ \bibinfo {pages} {054034}
  (\bibinfo {year} {2014})},\ \Eprint {http://arxiv.org/abs/1307.1896}
  {arXiv:1307.1896 [nucl-th]} \BibitemShut {NoStop}%
\bibitem [{\citenamefont {Fettes}\ and\ \citenamefont
  {Mei\ss{}ner}(2001)}]{Fettes:2000bb}%
  \BibitemOpen
  \bibfield  {author} {\bibinfo {author} {\bibfnamefont {Nadia}\ \bibnamefont
  {Fettes}}\ and\ \bibinfo {author} {\bibfnamefont {Ulf-G.}\ \bibnamefont
  {Mei\ss{}ner}},\ }\bibfield  {title} {\enquote {\bibinfo {title} {{Pion -
  nucleon scattering in an effective chiral field theory with explicit spin-3/2
  fields}},}\ }\href {\doibase 10.1016/S0375-9474(00)00368-7} {\bibfield
  {journal} {\bibinfo  {journal} {Nucl. Phys. A}\ }\textbf {\bibinfo {volume}
  {679}},\ \bibinfo {pages} {629--670} (\bibinfo {year} {2001})},\ \Eprint
  {http://arxiv.org/abs/hep-ph/0006299} {arXiv:hep-ph/0006299 [hep-ph]}
  \BibitemShut {NoStop}%
\bibitem [{\citenamefont {Hildebrandt}(2005)}]{Hildebrandt:2005ix}%
  \BibitemOpen
  \bibfield  {author} {\bibinfo {author} {\bibfnamefont {Robert~P.}\
  \bibnamefont {Hildebrandt}},\ }\emph {\bibinfo {title} {{Elastic Compton
  scattering from the nucleon and deuteron}}},\ \href@noop {} {Ph.D. thesis},\
  \bibinfo  {school} {Munich, Tech. U.} (\bibinfo {year} {2005}),\ \Eprint
  {http://arxiv.org/abs/nucl-th/0512064} {arXiv:nucl-th/0512064 [nucl-th]}
  \BibitemShut {NoStop}%
\bibitem [{\citenamefont {Bernard}\ \emph {et~al.}(2005)\citenamefont
  {Bernard}, \citenamefont {Hemmert},\ and\ \citenamefont
  {Meissner}}]{Bernard:2005fy}%
  \BibitemOpen
  \bibfield  {author} {\bibinfo {author} {\bibfnamefont {Veronique}\
  \bibnamefont {Bernard}}, \bibinfo {author} {\bibfnamefont {Thomas~R.}\
  \bibnamefont {Hemmert}}, \ and\ \bibinfo {author} {\bibfnamefont {Ulf-G.}\
  \bibnamefont {Meissner}},\ }\bibfield  {title} {\enquote {\bibinfo {title}
  {{Chiral extrapolations and the covariant small scale expansion}},}\ }\href
  {\doibase 10.1016/j.physletb.2005.06.088} {\bibfield  {journal} {\bibinfo
  {journal} {Phys. Lett. B}\ }\textbf {\bibinfo {volume} {622}},\ \bibinfo
  {pages} {141--150} (\bibinfo {year} {2005})},\ \Eprint
  {http://arxiv.org/abs/hep-lat/0503022} {arXiv:hep-lat/0503022 [hep-lat]}
  \BibitemShut {NoStop}%
\bibitem [{\citenamefont {Pascalutsa}\ and\ \citenamefont
  {Vanderhaeghen}(2006)}]{Pascalutsa:2005vq}%
  \BibitemOpen
  \bibfield  {author} {\bibinfo {author} {\bibfnamefont {Vladimir}\
  \bibnamefont {Pascalutsa}}\ and\ \bibinfo {author} {\bibfnamefont {Marc}\
  \bibnamefont {Vanderhaeghen}},\ }\bibfield  {title} {\enquote {\bibinfo
  {title} {{Chiral effective-field theory in the $\Delta$(1232) region: Pion
  electroproduction on the nucleon}},}\ }\href {\doibase
  10.1103/PhysRevD.73.034003} {\bibfield  {journal} {\bibinfo  {journal} {Phys.
  Rev. D}\ }\textbf {\bibinfo {volume} {73}},\ \bibinfo {pages} {034003}
  (\bibinfo {year} {2006})},\ \Eprint {http://arxiv.org/abs/hep-ph/0512244}
  {arXiv:hep-ph/0512244 [hep-ph]} \BibitemShut {NoStop}%
\bibitem [{\citenamefont {Liu}\ and\ \citenamefont {Zhu}(2007)}]{Liu:2007ct}%
  \BibitemOpen
  \bibfield  {author} {\bibinfo {author} {\bibfnamefont {Yan-Rui}\ \bibnamefont
  {Liu}}\ and\ \bibinfo {author} {\bibfnamefont {Shi-Lin}\ \bibnamefont
  {Zhu}},\ }\bibfield  {title} {\enquote {\bibinfo {title} {{Decuplet
  contribution to the meson-baryon scattering lengths}},}\ }\href {\doibase
  10.1140/epjc/s10052-007-0348-x} {\bibfield  {journal} {\bibinfo  {journal}
  {Eur. Phys. J.}\ }\textbf {\bibinfo {volume} {C52}},\ \bibinfo {pages}
  {177--186} (\bibinfo {year} {2007})},\ \Eprint
  {http://arxiv.org/abs/hep-ph/0702246} {arXiv:hep-ph/0702246 [HEP-PH]}
  \BibitemShut {NoStop}%
\bibitem [{\citenamefont {Liu}\ \emph {et~al.}(2011)\citenamefont {Liu},
  \citenamefont {Liu},\ and\ \citenamefont {Zhu}}]{Liu:2010bw}%
  \BibitemOpen
  \bibfield  {author} {\bibinfo {author} {\bibfnamefont {Zhan-Wei}\
  \bibnamefont {Liu}}, \bibinfo {author} {\bibfnamefont {Yan-Rui}\ \bibnamefont
  {Liu}}, \ and\ \bibinfo {author} {\bibfnamefont {Shi-Lin}\ \bibnamefont
  {Zhu}},\ }\bibfield  {title} {\enquote {\bibinfo {title} {{Pseudoscalar Meson
  and Decuplet Baryon Scattering Lengths}},}\ }\href {\doibase
  10.1103/PhysRevD.83.034004} {\bibfield  {journal} {\bibinfo  {journal} {Phys.
  Rev.}\ }\textbf {\bibinfo {volume} {D83}},\ \bibinfo {pages} {034004}
  (\bibinfo {year} {2011})},\ \Eprint {http://arxiv.org/abs/1011.3613}
  {arXiv:1011.3613 [hep-ph]} \BibitemShut {NoStop}%
\bibitem [{\citenamefont {Alarc\'{o}n}\ \emph {et~al.}(2013)\citenamefont
  {Alarc\'{o}n}, \citenamefont {Martin~Camalich},\ and\ \citenamefont
  {Oller}}]{Alarcon:2012kn}%
  \BibitemOpen
  \bibfield  {author} {\bibinfo {author} {\bibfnamefont {J.~M.}\ \bibnamefont
  {Alarc\'{o}n}}, \bibinfo {author} {\bibfnamefont {J.}~\bibnamefont
  {Martin~Camalich}}, \ and\ \bibinfo {author} {\bibfnamefont {J.~A.}\
  \bibnamefont {Oller}},\ }\bibfield  {title} {\enquote {\bibinfo {title}
  {{Improved description of the $\pi N$-scattering phenomenology in covariant
  baryon chiral perturbation theory}},}\ }\href {\doibase
  10.1016/j.aop.2013.06.001} {\bibfield  {journal} {\bibinfo  {journal} {Ann.
  Phys. (NY)}\ }\textbf {\bibinfo {volume} {336}},\ \bibinfo {pages} {413--461}
  (\bibinfo {year} {2013})},\ \Eprint {http://arxiv.org/abs/1210.4450}
  {arXiv:1210.4450 [hep-ph]} \BibitemShut {NoStop}%
\bibitem [{\citenamefont {McGovern}\ \emph {et~al.}(2013)\citenamefont
  {McGovern}, \citenamefont {Phillips},\ and\ \citenamefont
  {Grie\ss{}hammer}}]{McGovern:2012ew}%
  \BibitemOpen
  \bibfield  {author} {\bibinfo {author} {\bibfnamefont {J.~A.}\ \bibnamefont
  {McGovern}}, \bibinfo {author} {\bibfnamefont {D.~R.}\ \bibnamefont
  {Phillips}}, \ and\ \bibinfo {author} {\bibfnamefont {H.~W.}\ \bibnamefont
  {Grie\ss{}hammer}},\ }\bibfield  {title} {\enquote {\bibinfo {title}
  {{Compton scattering from the proton in an effective field theory with
  explicit Delta degrees of freedom}},}\ }\href {\doibase
  10.1140/epja/i2013-13012-1} {\bibfield  {journal} {\bibinfo  {journal} {Eur.
  Phys. J. A}\ }\textbf {\bibinfo {volume} {49}},\ \bibinfo {pages} {12}
  (\bibinfo {year} {2013})},\ \Eprint {http://arxiv.org/abs/1210.4104}
  {arXiv:1210.4104 [nucl-th]} \BibitemShut {NoStop}%
\bibitem [{\citenamefont {Yao}\ \emph {et~al.}(2016)\citenamefont {Yao},
  \citenamefont {Siemens}, \citenamefont {Bernard}, \citenamefont {Epelbaum},
  \citenamefont {Gasparyan}, \citenamefont {Gegelia}, \citenamefont {Krebs},\
  and\ \citenamefont {Mei\ss{}ner}}]{Yao:2016vbz}%
  \BibitemOpen
  \bibfield  {author} {\bibinfo {author} {\bibfnamefont {De-Liang}\
  \bibnamefont {Yao}}, \bibinfo {author} {\bibfnamefont {D.}~\bibnamefont
  {Siemens}}, \bibinfo {author} {\bibfnamefont {V.}~\bibnamefont {Bernard}},
  \bibinfo {author} {\bibfnamefont {E.}~\bibnamefont {Epelbaum}}, \bibinfo
  {author} {\bibfnamefont {A.~M.}\ \bibnamefont {Gasparyan}}, \bibinfo {author}
  {\bibfnamefont {J.}~\bibnamefont {Gegelia}}, \bibinfo {author} {\bibfnamefont
  {H.}~\bibnamefont {Krebs}}, \ and\ \bibinfo {author} {\bibfnamefont {Ulf-G.}\
  \bibnamefont {Mei\ss{}ner}},\ }\bibfield  {title} {\enquote {\bibinfo {title}
  {{Pion-nucleon scattering in covariant baryon chiral perturbation theory with
  explicit Delta resonances}},}\ }\href {\doibase 10.1007/JHEP05(2016)038}
  {\bibfield  {journal} {\bibinfo  {journal} {Journal of High Energy Physics}\
  }\textbf {\bibinfo {volume} {05}},\ \bibinfo {pages} {038} (\bibinfo {year}
  {2016})},\ \Eprint {http://arxiv.org/abs/1603.03638} {arXiv:1603.03638
  [hep-ph]} \BibitemShut {NoStop}%
\bibitem [{\citenamefont {Hiller~Blin}(2016)}]{HillerBlin:2016jpb}%
  \BibitemOpen
  \bibfield  {author} {\bibinfo {author} {\bibfnamefont {Astrid~Nathalie}\
  \bibnamefont {Hiller~Blin}},\ }\emph {\bibinfo {title} {{Electromagnetic
  interactions of light hadrons in covariant chiral perturbation theory}}},\
  \href@noop {} {Ph.D. thesis},\ \bibinfo  {school} {Valencia U., IFIC}
  (\bibinfo {year} {2016})\BibitemShut {NoStop}%
\bibitem [{\citenamefont {Dirac}(1936)}]{Dirac:1936tg}%
  \BibitemOpen
  \bibfield  {author} {\bibinfo {author} {\bibfnamefont {P.~A.~M.}\
  \bibnamefont {Dirac}},\ }\bibfield  {title} {\enquote {\bibinfo {title}
  {{Relativistic wave equations}},}\ }\href {\doibase 10.1098/rspa.1936.0111}
  {\bibfield  {journal} {\bibinfo  {journal} {Proc. R. Soc. Lond. A}\ }\textbf
  {\bibinfo {volume} {155}},\ \bibinfo {pages} {447--459} (\bibinfo {year}
  {1936})}\BibitemShut {NoStop}%
\bibitem [{\citenamefont {Fierz}\ and\ \citenamefont
  {Pauli}(1939)}]{Fierz:1939ix}%
  \BibitemOpen
  \bibfield  {author} {\bibinfo {author} {\bibfnamefont {M.}~\bibnamefont
  {Fierz}}\ and\ \bibinfo {author} {\bibfnamefont {W.}~\bibnamefont {Pauli}},\
  }\bibfield  {title} {\enquote {\bibinfo {title} {{On relativistic wave
  equations for particles of arbitrary spin in an electromagnetic field}},}\
  }\href {\doibase 10.1098/rspa.1939.0140} {\bibfield  {journal} {\bibinfo
  {journal} {Proc. R. Soc. Lond. A}\ }\textbf {\bibinfo {volume} {A173}},\
  \bibinfo {pages} {211--232} (\bibinfo {year} {1939})}\BibitemShut {NoStop}%
\bibitem [{\citenamefont {Rarita}\ and\ \citenamefont
  {Schwinger}(1941)}]{Rarita:1941mf}%
  \BibitemOpen
  \bibfield  {author} {\bibinfo {author} {\bibfnamefont {William}\ \bibnamefont
  {Rarita}}\ and\ \bibinfo {author} {\bibfnamefont {Julian}\ \bibnamefont
  {Schwinger}},\ }\bibfield  {title} {\enquote {\bibinfo {title} {{On a theory
  of particles with half integral spin}},}\ }\href {\doibase
  10.1103/PhysRev.60.61} {\bibfield  {journal} {\bibinfo  {journal} {Phys.
  Rev.}\ }\textbf {\bibinfo {volume} {60}},\ \bibinfo {pages} {61} (\bibinfo
  {year} {1941})}\BibitemShut {NoStop}%
\bibitem [{\citenamefont {{Gupta}}(1954)}]{1954PhRv...95.1334G}%
  \BibitemOpen
  \bibfield  {author} {\bibinfo {author} {\bibfnamefont {Suraj~N.}\
  \bibnamefont {{Gupta}}},\ }\bibfield  {title} {\enquote {\bibinfo {title}
  {{Fierz-Pauli theory of particles of spin 3/2}},}\ }\href {\doibase
  10.1103/PhysRev.95.1334} {\bibfield  {journal} {\bibinfo  {journal} {Physical
  Review}\ }\textbf {\bibinfo {volume} {95}},\ \bibinfo {pages} {1334--1341}
  (\bibinfo {year} {1954})}\BibitemShut {NoStop}%
\bibitem [{\citenamefont {Moldauer}\ and\ \citenamefont
  {Case}(1956)}]{Moldauer:1956zz}%
  \BibitemOpen
  \bibfield  {author} {\bibinfo {author} {\bibfnamefont {P.~A.}\ \bibnamefont
  {Moldauer}}\ and\ \bibinfo {author} {\bibfnamefont {K.~M.}\ \bibnamefont
  {Case}},\ }\bibfield  {title} {\enquote {\bibinfo {title} {{Properties of
  half-integral spin Dirac-Fierz-Pauli particles}},}\ }\href {\doibase
  10.1103/PhysRev.102.279} {\bibfield  {journal} {\bibinfo  {journal} {Phys.
  Rev.}\ }\textbf {\bibinfo {volume} {102}},\ \bibinfo {pages} {279--285}
  (\bibinfo {year} {1956})}\BibitemShut {NoStop}%
\bibitem [{\citenamefont {Fronsdal}(1958)}]{1958NCim....9S.416F}%
  \BibitemOpen
  \bibfield  {author} {\bibinfo {author} {\bibfnamefont {C.}~\bibnamefont
  {Fronsdal}},\ }\bibfield  {title} {\enquote {\bibinfo {title} {{On the theory
  of higher spin fields}},}\ }\href {\doibase 10.1007/BF02747684} {\bibfield
  {journal} {\bibinfo  {journal} {Il Nuovo Cimento}\ }\textbf {\bibinfo
  {volume} {9}},\ \bibinfo {pages} {416--443} (\bibinfo {year}
  {1958})}\BibitemShut {NoStop}%
\bibitem [{\citenamefont {Aurilia}\ and\ \citenamefont
  {Umezawa}(1969)}]{Aurilia:1969bg}%
  \BibitemOpen
  \bibfield  {author} {\bibinfo {author} {\bibfnamefont {Antonio}\ \bibnamefont
  {Aurilia}}\ and\ \bibinfo {author} {\bibfnamefont {H.}~\bibnamefont
  {Umezawa}},\ }\bibfield  {title} {\enquote {\bibinfo {title} {{Theory of
  high-spin fields}},}\ }\href {\doibase 10.1103/PhysRev.182.1682} {\bibfield
  {journal} {\bibinfo  {journal} {Phys. Rev.}\ }\textbf {\bibinfo {volume}
  {182}},\ \bibinfo {pages} {1682--1694} (\bibinfo {year} {1969})}\BibitemShut
  {NoStop}%
\bibitem [{\citenamefont {Van~Nieuwenhuizen}(1981)}]{VanNieuwenhuizen:1981ae}%
  \BibitemOpen
  \bibfield  {author} {\bibinfo {author} {\bibfnamefont {P.}~\bibnamefont
  {Van~Nieuwenhuizen}},\ }\bibfield  {title} {\enquote {\bibinfo {title}
  {{Supergravity}},}\ }\href {\doibase 10.1016/0370-1573(81)90157-5} {\bibfield
   {journal} {\bibinfo  {journal} {Phys. Rept.}\ }\textbf {\bibinfo {volume}
  {68}},\ \bibinfo {pages} {189--398} (\bibinfo {year} {1981})}\BibitemShut
  {NoStop}%
\bibitem [{\citenamefont {Williams}(1985)}]{Williams:1985zz}%
  \BibitemOpen
  \bibfield  {author} {\bibinfo {author} {\bibfnamefont {H.~T.}\ \bibnamefont
  {Williams}},\ }\bibfield  {title} {\enquote {\bibinfo {title}
  {{Misconceptions regarding spin 3/2}},}\ }\href {\doibase
  10.1103/PhysRevC.31.2297} {\bibfield  {journal} {\bibinfo  {journal} {Phys.
  Rev. C}\ }\textbf {\bibinfo {volume} {31}},\ \bibinfo {pages} {2297--2299}
  (\bibinfo {year} {1985})}\BibitemShut {NoStop}%
\bibitem [{\citenamefont {Benmerrouche}\ \emph {et~al.}(1989)\citenamefont
  {Benmerrouche}, \citenamefont {Davidson},\ and\ \citenamefont
  {Mukhopadhyay}}]{Benmerrouche:1989uc}%
  \BibitemOpen
  \bibfield  {author} {\bibinfo {author} {\bibfnamefont {M.}~\bibnamefont
  {Benmerrouche}}, \bibinfo {author} {\bibfnamefont {R.~M.}\ \bibnamefont
  {Davidson}}, \ and\ \bibinfo {author} {\bibfnamefont {Nimai~C.}\ \bibnamefont
  {Mukhopadhyay}},\ }\bibfield  {title} {\enquote {\bibinfo {title} {{Problems
  of describing spin 3/2 baryon resonances in the effective Lagrangian
  theory}},}\ }\href {\doibase 10.1103/PhysRevC.39.2339} {\bibfield  {journal}
  {\bibinfo  {journal} {Phys. Rev. C}\ }\textbf {\bibinfo {volume} {39}},\
  \bibinfo {pages} {2339--2348} (\bibinfo {year} {1989})}\BibitemShut {NoStop}%
\bibitem [{\citenamefont {Pascalutsa}(1994)}]{Pascalutsa:1994tp}%
  \BibitemOpen
  \bibfield  {author} {\bibinfo {author} {\bibfnamefont {Vladimir}\
  \bibnamefont {Pascalutsa}},\ }\bibfield  {title} {\enquote {\bibinfo {title}
  {{On the interaction of spin 3/2 particles}},}\ }\href@noop {} {\  (\bibinfo
  {year} {1994})},\ \Eprint {http://arxiv.org/abs/hep-ph/9412321}
  {arXiv:hep-ph/9412321 [hep-ph]} \BibitemShut {NoStop}%
\bibitem [{\citenamefont {Haberzettl}(1998)}]{Haberzettl:1998rw}%
  \BibitemOpen
  \bibfield  {author} {\bibinfo {author} {\bibfnamefont {Helmut}\ \bibnamefont
  {Haberzettl}},\ }\bibfield  {title} {\enquote {\bibinfo {title} {{Propagation
  of a massive spin-3/2 particle}},}\ }\href@noop {} {\  (\bibinfo {year}
  {1998})},\ \Eprint {http://arxiv.org/abs/nucl-th/9812043}
  {arXiv:nucl-th/9812043 [nucl-th]} \BibitemShut {NoStop}%
\bibitem [{\citenamefont {Pilling}(2005)}]{Pilling:2004cu}%
  \BibitemOpen
  \bibfield  {author} {\bibinfo {author} {\bibfnamefont {Terry}\ \bibnamefont
  {Pilling}},\ }\bibfield  {title} {\enquote {\bibinfo {title} {{Symmetry of
  massive Rarita-Schwinger fields}},}\ }\href {\doibase
  10.1142/S0217751X05021300} {\bibfield  {journal} {\bibinfo  {journal} {Int.
  J. Mod. Phys. A}\ }\textbf {\bibinfo {volume} {20}},\ \bibinfo {pages}
  {2715--2741} (\bibinfo {year} {2005})},\ \Eprint
  {http://arxiv.org/abs/hep-th/0404131} {arXiv:hep-th/0404131 [hep-th]}
  \BibitemShut {NoStop}%
\bibitem [{\citenamefont {{Kaloshin}}\ and\ \citenamefont
  {{Lomov}}(2006)}]{2006PAN....69..541K}%
  \BibitemOpen
  \bibfield  {author} {\bibinfo {author} {\bibfnamefont {A.~E.}\ \bibnamefont
  {{Kaloshin}}}\ and\ \bibinfo {author} {\bibfnamefont {V.~P.}\ \bibnamefont
  {{Lomov}}},\ }\bibfield  {title} {\enquote {\bibinfo {title}
  {{Rarita-Schwinger field: Dressing procedure and spin-parity of
  components}},}\ }\href {\doibase 10.1134/S1063778806030161} {\bibfield
  {journal} {\bibinfo  {journal} {Physics of Atomic Nuclei}\ }\textbf {\bibinfo
  {volume} {69}},\ \bibinfo {pages} {541--551} (\bibinfo {year} {2006})},\
  \Eprint {http://arxiv.org/abs/hep-ph/0409052} {hep-ph/0409052} \BibitemShut
  {NoStop}%
\bibitem [{\citenamefont {Tang}\ and\ \citenamefont
  {Ellis}(1996)}]{Tang:1996sq}%
  \BibitemOpen
  \bibfield  {author} {\bibinfo {author} {\bibfnamefont {Hua-Bin}\ \bibnamefont
  {Tang}}\ and\ \bibinfo {author} {\bibfnamefont {Paul~J.}\ \bibnamefont
  {Ellis}},\ }\bibfield  {title} {\enquote {\bibinfo {title} {{Redundance of
  Delta isobar parameters in effective field theories}},}\ }\href {\doibase
  10.1016/0370-2693(96)00862-3} {\bibfield  {journal} {\bibinfo  {journal}
  {Phys. Lett. B}\ }\textbf {\bibinfo {volume} {387}},\ \bibinfo {pages}
  {9--13} (\bibinfo {year} {1996})},\ \Eprint
  {http://arxiv.org/abs/hep-ph/9606432} {arXiv:hep-ph/9606432 [hep-ph]}
  \BibitemShut {NoStop}%
\bibitem [{\citenamefont {Hacker}\ \emph {et~al.}(2005)\citenamefont {Hacker},
  \citenamefont {Wies}, \citenamefont {Gegelia},\ and\ \citenamefont
  {Scherer}}]{Hacker:2005fh}%
  \BibitemOpen
  \bibfield  {author} {\bibinfo {author} {\bibfnamefont {C.}~\bibnamefont
  {Hacker}}, \bibinfo {author} {\bibfnamefont {N.}~\bibnamefont {Wies}},
  \bibinfo {author} {\bibfnamefont {J.}~\bibnamefont {Gegelia}}, \ and\
  \bibinfo {author} {\bibfnamefont {S.}~\bibnamefont {Scherer}},\ }\bibfield
  {title} {\enquote {\bibinfo {title} {{Including the $\Delta$(1232) resonance
  in baryon chiral perturbation theory}},}\ }\href {\doibase
  10.1103/PhysRevC.72.055203} {\bibfield  {journal} {\bibinfo  {journal} {Phys.
  Rev.}\ }\textbf {\bibinfo {volume} {C72}},\ \bibinfo {pages} {055203}
  (\bibinfo {year} {2005})},\ \Eprint {http://arxiv.org/abs/hep-ph/0505043}
  {arXiv:hep-ph/0505043 [hep-ph]} \BibitemShut {NoStop}%
\bibitem [{\citenamefont {Krebs}\ \emph {et~al.}(2010)\citenamefont {Krebs},
  \citenamefont {Epelbaum},\ and\ \citenamefont {Mei\ss{}ner}}]{Krebs:2009bf}%
  \BibitemOpen
  \bibfield  {author} {\bibinfo {author} {\bibfnamefont {H.}~\bibnamefont
  {Krebs}}, \bibinfo {author} {\bibfnamefont {E.}~\bibnamefont {Epelbaum}}, \
  and\ \bibinfo {author} {\bibfnamefont {U.~G.}\ \bibnamefont {Mei\ss{}ner}},\
  }\bibfield  {title} {\enquote {\bibinfo {title} {{Redundancy of the off-shell
  parameters in chiral effective field theory with explicit spin-3/2 degrees of
  freedom}},}\ }\href {\doibase 10.1016/j.physletb.2009.12.023} {\bibfield
  {journal} {\bibinfo  {journal} {Phys. Lett. B}\ }\textbf {\bibinfo {volume}
  {683}},\ \bibinfo {pages} {222--228} (\bibinfo {year} {2010})},\ \Eprint
  {http://arxiv.org/abs/0905.2744} {arXiv:0905.2744 [hep-th]} \BibitemShut
  {NoStop}%
\bibitem [{\citenamefont {Scherer}\ and\ \citenamefont
  {Schindler}(2012)}]{Scherer:2012xha}%
  \BibitemOpen
  \bibfield  {author} {\bibinfo {author} {\bibfnamefont {Stefan}\ \bibnamefont
  {Scherer}}\ and\ \bibinfo {author} {\bibfnamefont {Matthias~R.}\ \bibnamefont
  {Schindler}},\ }\href {\doibase 10.1007/978-3-642-19254-8} {\emph {\bibinfo
  {title} {A Primer for Chiral Perturbation Theory}}},\ Vol.\ \bibinfo {volume}
  {830}\ (\bibinfo  {publisher} {Springer-Verlag Berlin Heidelberg},\ \bibinfo
  {year} {2012})\ pp.\ \bibinfo {pages} {1--338}\BibitemShut {NoStop}%
\bibitem [{\citenamefont {Kamefuchi}\ \emph {et~al.}(1961)\citenamefont
  {Kamefuchi}, \citenamefont {O'Raifeartaigh},\ and\ \citenamefont
  {Salam}}]{Kamefuchi:1961sb}%
  \BibitemOpen
  \bibfield  {author} {\bibinfo {author} {\bibfnamefont {S.}~\bibnamefont
  {Kamefuchi}}, \bibinfo {author} {\bibfnamefont {L.}~\bibnamefont
  {O'Raifeartaigh}}, \ and\ \bibinfo {author} {\bibfnamefont {Abdus}\
  \bibnamefont {Salam}},\ }\bibfield  {title} {\enquote {\bibinfo {title}
  {{Change of variables and equivalence theorems in quantum field theories}},}\
  }\href {\doibase 10.1016/0029-5582(61)90056-6,10.1016/0029-5582(61)91075-6}
  {\bibfield  {journal} {\bibinfo  {journal} {Nucl. Phys.}\ }\textbf {\bibinfo
  {volume} {28}},\ \bibinfo {pages} {529--549} (\bibinfo {year}
  {1961})}\BibitemShut {NoStop}%
\bibitem [{\citenamefont {Nath}\ \emph {et~al.}(1971)\citenamefont {Nath},
  \citenamefont {Etemadi},\ and\ \citenamefont {Kimel}}]{Nath:1971wp}%
  \BibitemOpen
  \bibfield  {author} {\bibinfo {author} {\bibfnamefont {L.~M.}\ \bibnamefont
  {Nath}}, \bibinfo {author} {\bibfnamefont {B.}~\bibnamefont {Etemadi}}, \
  and\ \bibinfo {author} {\bibfnamefont {J.~D.}\ \bibnamefont {Kimel}},\
  }\bibfield  {title} {\enquote {\bibinfo {title} {{Uniqueness of the
  interaction involving spin 3/2 particles}},}\ }\href {\doibase
  10.1103/PhysRevD.3.2153} {\bibfield  {journal} {\bibinfo  {journal} {Phys.
  Rev. D}\ }\textbf {\bibinfo {volume} {3}},\ \bibinfo {pages} {2153--2161}
  (\bibinfo {year} {1971})}\BibitemShut {NoStop}%
\bibitem [{\citenamefont {Hemmert}\ \emph {et~al.}(1997)\citenamefont
  {Hemmert}, \citenamefont {Holstein},\ and\ \citenamefont
  {Kambor}}]{Hemmert:1996xg}%
  \BibitemOpen
  \bibfield  {author} {\bibinfo {author} {\bibfnamefont {Thomas~R.}\
  \bibnamefont {Hemmert}}, \bibinfo {author} {\bibfnamefont {Barry~R.}\
  \bibnamefont {Holstein}}, \ and\ \bibinfo {author} {\bibfnamefont {Joachim}\
  \bibnamefont {Kambor}},\ }\bibfield  {title} {\enquote {\bibinfo {title}
  {{Systematic $1/M$ expansion for spin 3/2 particles in baryon chiral
  perturbation theory}},}\ }\href {\doibase 10.1016/S0370-2693(97)00049-X}
  {\bibfield  {journal} {\bibinfo  {journal} {Phys. Lett. B}\ }\textbf
  {\bibinfo {volume} {395}},\ \bibinfo {pages} {89--95} (\bibinfo {year}
  {1997})},\ \Eprint {http://arxiv.org/abs/hep-ph/9606456}
  {arXiv:hep-ph/9606456 [hep-ph]} \BibitemShut {NoStop}%
\bibitem [{\citenamefont {Hemmert}(1997)}]{Hemmert:1997wz}%
  \BibitemOpen
  \bibfield  {author} {\bibinfo {author} {\bibfnamefont {Thomas~R.}\
  \bibnamefont {Hemmert}},\ }\emph {\bibinfo {title} {{Heavy baryon chiral
  perturbation theory with light deltas}}},\ \href@noop {} {Ph.D. thesis},\
  \bibinfo  {school} {Massachusetts U., Amherst} (\bibinfo {year}
  {1997})\BibitemShut {NoStop}%
\bibitem [{\citenamefont {Krebs}\ \emph {et~al.}(2009)\citenamefont {Krebs},
  \citenamefont {Epelbaum},\ and\ \citenamefont {Meissner}}]{Krebs:2008zb}%
  \BibitemOpen
  \bibfield  {author} {\bibinfo {author} {\bibfnamefont {H.}~\bibnamefont
  {Krebs}}, \bibinfo {author} {\bibfnamefont {E.}~\bibnamefont {Epelbaum}}, \
  and\ \bibinfo {author} {\bibfnamefont {Ulf-G.}\ \bibnamefont {Meissner}},\
  }\bibfield  {title} {\enquote {\bibinfo {title} {{On-shell consistency of the
  Rarita-Schwinger field formulation}},}\ }\href {\doibase
  10.1103/PhysRevC.80.028201} {\bibfield  {journal} {\bibinfo  {journal} {Phys.
  Rev.}\ }\textbf {\bibinfo {volume} {C80}},\ \bibinfo {pages} {028201}
  (\bibinfo {year} {2009})},\ \Eprint {http://arxiv.org/abs/0812.0132}
  {arXiv:0812.0132 [hep-th]} \BibitemShut {NoStop}%
\bibitem [{\citenamefont {Wies}\ \emph {et~al.}(2006)\citenamefont {Wies},
  \citenamefont {Gegelia},\ and\ \citenamefont {Scherer}}]{Wies:2006rv}%
  \BibitemOpen
  \bibfield  {author} {\bibinfo {author} {\bibfnamefont {N.}~\bibnamefont
  {Wies}}, \bibinfo {author} {\bibfnamefont {J.}~\bibnamefont {Gegelia}}, \
  and\ \bibinfo {author} {\bibfnamefont {S.}~\bibnamefont {Scherer}},\
  }\bibfield  {title} {\enquote {\bibinfo {title} {{Consistency of the
  $\pi\Delta$ interaction in chiral perturbation theory}},}\ }\href {\doibase
  10.1103/PhysRevD.73.094012} {\bibfield  {journal} {\bibinfo  {journal} {Phys.
  Rev. D}\ }\textbf {\bibinfo {volume} {73}},\ \bibinfo {pages} {094012}
  (\bibinfo {year} {2006})},\ \Eprint {http://arxiv.org/abs/hep-ph/0602073}
  {arXiv:hep-ph/0602073 [hep-ph]} \BibitemShut {NoStop}%
\bibitem [{\citenamefont {Long}\ and\ \citenamefont
  {Lensky}(2011)}]{Long:2010kt}%
  \BibitemOpen
  \bibfield  {author} {\bibinfo {author} {\bibfnamefont {Bingwei}\ \bibnamefont
  {Long}}\ and\ \bibinfo {author} {\bibfnamefont {Vadim}\ \bibnamefont
  {Lensky}},\ }\bibfield  {title} {\enquote {\bibinfo {title} {{Heavy-particle
  formalism with Foldy-Wouthuysen representation}},}\ }\href {\doibase
  10.1103/PhysRevC.83.045206} {\bibfield  {journal} {\bibinfo  {journal} {Phys.
  Rev. C}\ }\textbf {\bibinfo {volume} {83}},\ \bibinfo {pages} {045206}
  (\bibinfo {year} {2011})},\ \Eprint {http://arxiv.org/abs/1010.2738}
  {arXiv:1010.2738} \BibitemShut {NoStop}%
\end{thebibliography}%
\end{document}